\documentclass[useAMS,usenatbib]{mn2e}
\usepackage{txfonts,graphicx}
\usepackage{natbib,longtable,url}

\newcommand{\Mn}[5]{\mbox{$#1\,^#2{\rm #3}^{{\rm #4}}_{\rm #5}$}}
\newcommand{\Co}[5]{\mbox{$#1\,^#2{\rm #3}^{{\rm #4}}_{\rm #5}$}}

\newcommand\ion[2]{#1$\;${\scshape{#2}}} %ion, i.e., MnII = \ion{Mn}{ii}
\newcommand{\opd}{\log \tau_{\rm 5000}}
\newcommand{\SH}{S\!_{\rm H}}           % Hydrogen collision multiplier
\newcommand{\SP}{S\!_{\rm P}}           % Photoionization multiplier
\newcommand{\Te}{T_{\rm e}}             % Kinetic temperature
\newcommand{\mA}{{\rm m\AA}}            % milli angstrem - 10^-12 cm
\newcommand{\Elow}{E_{\rm low}}         % low energy
\newcommand{\loge}{\log\varepsilon}      % log epsilon

\newcommand{\Teff}{\ensuremath{T_{\mathrm{eff}}}}     % T_eff
\newcommand{\kms}{km s$^{-1}$}
\newcommand{\Vmic}{\xi_{\rm{t}}}                 % microturbulence velocity
\newcommand{\loggfestar}{\log (gf\varepsilon)_{\ast}}
\newcommand{\loggfesun}{\log (gf\varepsilon)_{\odot}}
\newcommand{\logeMnN}[1]{\log\varepsilon_{\rm Mn, \odot}^{\rm NLTE}}
\newcommand{\logeMnL}[1]{\log\varepsilon_{\rm Mn, \odot}^{\rm LTE}}
\newcommand{\logeCoN}[1]{\log\varepsilon_{\rm Co, \odot}^{\rm NLTE}}
\newcommand{\logeCoL}[1]{\log\varepsilon_{\rm Co, \odot}^{\rm LTE}}
\newcommand{\logeCo}[1]{\log\varepsilon_{\rm Co, \odot}}
\newcommand{\logeMn}[1]{\log\varepsilon_{\rm Mn, \odot}}
\newcommand{\logeFesun}[1]{\log\varepsilon_{\rm Fe, \odot}}
\newcommand{\logeFe}[1]{\log\varepsilon_{\rm Fe, \star}}
\newcommand{\logemean}[1]{\log\varepsilon_{\rm mean}}

\title[Cobalt lines in spectra of the sun and metal-poor stars]{NLTE analysis of
\ion{Co}{i}/\ion{Co}{ii} lines in spectra of cool stars with new laboratory
hyperfine splitting constants \thanks{Based on observations collected at the
European Southern Observatory,Chile, 67.D-0086A, and the Calar Alto Observatory,
Spain.}}
\author[M. Bergemann, J. C. Pickering, and T. Gehren]{Maria
Bergemann$^{1,3}$\thanks{E-mail: mbergema@mpa-garching.mpg.de (MB)}, Juliet C.
Pickering$^{2}$\thanks{E-mail: j.pickering@imperial.ac.uk (JCP)}, and Thomas
Gehren$^{3}$\thanks{E-mail: gehren@usm.lmu.de (TG)}\\
$^{1}$Max-Planck Institute for Astrophysics, Karl-Schwarzschild Str. 1, 85741,
Garching, Germany \\
$^{2}$Physics Department, Blackett Laboratory, Imperial College, London SW7 2BZ,
UK \\
$^{3}$University Observatory, Ludwig-Maximilian University, Scheinerstr. 1, 
81679 Munich, Germany}

\begin{document}

\date{Accepted Date. Received Date; in original Date}

\pagerange{\pageref{firstpage}--\pageref{lastpage}} \pubyear{2009}

\maketitle

\label{firstpage}

\begin{abstract}
The analysis of stellar abundances for odd-$Z$ Fe-peak elements requires
accurate NLTE modelling of spectral lines fully taking into account the
hyperfine structure splitting (HFS) of lines. Here, we investigate the
statistical equilibrium of Co in the atmospheres of cool stars, and the
influence of NLTE and HFS on the formation of Co lines and abundances.
Significant departures from LTE level populations are found for \ion{Co}{i},
also number densities of excited states in \ion{Co}{ii} differ from LTE at low
metallicity. The NLTE level populations are used to determine the abundance of
Co in solar photosphere, $\loge = 4.95 \pm 0.04$ dex, which is in agreement with
that in \ion{C}{i} meteorites within the combined uncertainties. The spectral
lines of \ion{Co}{i} were calculated using the results of recent measurements of
hyperfine interaction constants by UV Fourier transform spectrometry. For
\ion{Co}{ii}, the first laboratory measurements of hyperfine structure
splitting $A$ and $B$ factors were performed. These highly accurate $A$ factor
measurements (errors of the order of 3-7\%) allow, for the first time, reliable
modelling of \ion{Co}{ii} lines in the solar and stellar spectra and, thus, a
test of the \ion{Co}{i}/\ion{Co}{ii} ionization equilibrium in stellar
atmospheres. A differential abundance analysis of Co is carried out for $18$
stars in the metallicity range $-3.12 <$ [Fe/H] $< 0$. The abundances are
derived by method of spectrum synthesis. At low [Fe/H], NLTE abundance
corrections for \ion{Co}{i} lines are as large as $+0.6 \ldots +0.8$ dex. Thus,
LTE abundances of Co in metal-poor stars are severely underestimated. The
stellar NLTE abundances determined from the single UV line of \ion{Co}{ii} are
lower by $\sim 0.5 - 0.6$ dex. The discrepancy might be attributed to possible 
blends that have not been accounted for in the solar \ion{Co}{ii} line and its
erroneous oscillator strength. The increasing [Co/Fe] trend in metal-poor stars,
as calculated from the \ion{Co}{i} lines under NLTE, can be explained if Co is
overproduced relative to Fe in massive stars. The models of galactic chemical
evolution are wholly inadequate to describe this trend suggesting that the
problem is in SN yields.
\end{abstract}


\section{Introduction}

Cobalt is one of the most intriguing elements in studies of galactic chemical
evolution. There is neither an agreement on the overall abundance trend of Co
in the halo and thin/thick disk, nor a generally accepted scenario for
the nucleosynthetic production of cobalt.

Several investigations of Co abundances have been done in various metallicity
regimes and stellar populations. All of these abundance investigations were 
based on \ion{Co}{i} lines and local thermodynamic equilibrium (LTE) for the
line formation. According to the results of \citet{1991A&A...241..501G}, Co
follows the depletion of Fe down to [Fe/H] $\approx -2.5$. At lower
metallicities, \citet{1995AJ....109.2757M} observe a strong increase of cobalt
abundance with respect to iron by $\sim 0.5$ dex. \citet{2005A&A...441.1149D}
obtained a different trend for [Co/Fe], characterized by a steady increase
towards supersolar ratios of $\sim +0.2$ from [Fe/H] $= 0$ to $-0.8$. An excess
of cobalt with respect to iron at $-4 \leq$ [Fe/H] $\leq -2.5$ was reported by
\citet{2008ApJ...681.1524L}, \citet{2004A&A...416.1117C}, and
\citet{1991AJ....102..303R}. \citet{2003MNRAS.340..304R,2006MNRAS.367.1329R}
find supersolar and mildly subsolar Co abundances for the thick and thin disk
stars, respectively.

The \emph{increasing} [Co/Fe] ratio with decreasing metallicity was attributed
to the Co production in both types of supernovae (SNe) but with uncertain
relative contributions \citep{2005A&A...441.1149D}. However, this is hard to
understand from the point of view of stellar nucleosynthesis. Cobalt is an
odd-$Z$ element of the Fe-group and is produced in $\alpha$-rich freeze-out in
SNe. Thus, its abundance must be sensitive to the explosion entropy and neutron
excess available to the progenitor star in the explosive Si-burning phase. The
neutron enrichment is determined by the initial metal content of a star and by
the previous hydrostatic burning stages. Hence, in metal-poor environments we
expect to see an \emph{odd-even effect} characterized by a general suppression
of the stable odd-$Z$ nuclei relative to their stable even-$Z$ neighbours
\citep{1957RvMP...29..547B}. Still, this effect is in conflict with the
increasing [Co/Fe] trend in metal-poor stars.

Thus far, any attempt to describe the abundances of Co within the framework of
explosive nucleosynthesis in SNe failed. Models of galactic chemical evolution
(GCE), which use standard metallicity-dependent prescriptions for Co production
in SNe II \citep{1995ApJS..101..181W} and SNe Ia
\citep{1984ApJ...286..644N,1999ApJS..125..439I} predict a decreasing or nearly
flat [Co/Fe] trend
\citep{1995ApJS...98..617T,1998ApJ...496..155S,2000A&A...359..191G}. Models with
metallicity-independent SN yields give only a qualitatively similar behaviour,
however large discrepancies with spectroscopic abundances are seen either for
the thick disk \citep{2000A&A...359..191G} or for the halo
\citep{2004A&A...421..613F}.

There are several explanations for the disagreement between observed Co trends
and predictions of chemical evolution models. Deficiencies of the latter are
hidden among others in the assumed shape of the initial mass function (IMF). As
shown by \citet{1993ApJ...406..580W}, the shape of the IMF introduces a factor
of $2$ in uncertainty to the absolute yield value of an element. The physics
of supernova explosions is still poorly understood \citep{2007PrPNP..59...74T},
hence theoretical stellar yields for Co may be erroneous. It is also important
to note that the majority of GCE models assume a single gas phase for the
galactic ISM, and only a few consider multiphase ISM
\citep[e.g.][]{2009MNRAS.396..203S} that is crucial for a physically accurate
description of the galactic chemo-dynamical evolution.

On the other hand, a close examination of the 'observed' Co abundances in the
Sun and metal-poor stars reveals that all of them rely on several coarse
assumptions: LTE in the line formation, static 1D atmospheric models, and,
often, neglect of hyperfine structure splitting (HFS) of lines. The errors in
abundances caused by these approximations are hard to predict, especially
because effects of NLTE and temperature inhomogeneities are not independent
\citep{2005ARA&A..43..481A}.
Iron is the only representative of the Fe-group for which calculations of 1.5D
NLTE line formation with 3D convective model atmospheres have been performed.
NLTE analyses with 1D model atmospheres exist for a number of elements,
specifically for Fe \citep[see][and references therein]{2003A&A...407..691K},
Sc \citep{2008A&A...481..489Z}, and Mn \citep{2008A&A...492..823B}.
These and other studies prove that the NLTE abundance corrections can be
significant at low gravities and/or metallicities. Due to the overionization,
NLTE abundances of Mn in metal-poor stars are larger by $+0.1 \ldots +0.6$ dex
than those computed under LTE.

Knowledge of hyperfine structure (HFS) in odd-$Z$ iron-group elements is
essential in calculations of line formation and abundances. The reason is that
HFS is not merely a line broadening mechanism; it modifies absorption over the
entire line profile, especially for strong and saturated lines. Thus, neglect of
HFS results in wavelength shifts, incorrectly determined damping parameters, and
erroneous abundances \citep[this last issue is addressed
in][]{1971A&A....10..434H,2000ApJ...537L..57P,2005A&A...441.1149D}. In a
strictly \emph{differential} analysis of stars with respect to the Sun, which is
usually preferred in order to minimize other deficiencies of the
stellar atmosphere modelling, neglect of HFS for the reference solar spectral
lines may lead to systematically \textit{underestimated} abundances in
metal-poor stars. The accuracy of the adopted HFS factors is also important, as
we will show in this paper. At present, high resolution measurements of HFS over
a wide spectral range are possible with high resolution Fourier transform
spectrometry (FTS). For example, an analysis of \ion{Co}{i} was undertaken
using over one thousand line profiles acquired with the FT spectrometer at
Imperial College, and yielded HFS $A$ splitting factors for $297$ energy levels,
almost all known \ion{Co}{i} energy levels \citep{1996ApJS..107..811P}.

In this paper, we investigate the excitation-ionization equilibrium of Co in
stellar atmospheres and predict what errors in abundance calculations occur
when NLTE effects and hyperfine splitting of lines are neglected. Given the 
particular importance of HFS in this context, we have measured hyperfine
splitting of \ion{Co}{ii} energy levels using the technique of FTS. We analyze
the abundance of Co in the Sun, and compare it with that of meteorites to
ascertain the reason for the discrepancies. Furthermore, we compute the NLTE
abundances for a sample of $18$ stars, which belong to the thin and thick disk,
and the halo. The evolution of abundance ratios in the Galaxy and some
implications for the Fe-peak element nucleosynthesis are discussed.
\section{Measurements of the hyperfine structure of \ion{Co}{ii}
levels}{\label{sec:meas}}
\subsection{Simple Theory}

For atoms with non zero nuclear spin {\em I} the fine structure levels undergo 
splitting because of the hyperfine interaction between the nucleus and the 
electrons. In the absence of perturbations the energy of the hyperfine structure
multiplets is given \citep{Kopfermann58} by:
\begin{equation}
W_{F}=W_{J} + \frac{1}{2}AK + B\frac{ (3/4)K(K+1) -J(J+1)I(I+1)}
{2I(2I-1)J(2J-1)}
\end{equation}
where $W_{J}$ is the energy of the fine structure level of quantum number $J$,
$A$ and $B$ are the magnetic dipole and the electric quadrupole hyperfine
interaction constants respectively, and $K$ is defined as
\[K=F(F+1)-J(J+1)-I(I+1) \] in which $F$ is the quantum number associated with
the total angular momentum of the atom,
\[F=I+J; \;\; I+J-1; \; ... ; \; |I-J| \]
The selection rules which govern the hyperfine transitions are:
$ \Delta F = 0; \pm 1 \;\;\;\; $but not $ \;\;F=0 \leftrightarrow F=0 $

In addition there are intensity rules \citep{Kuhn62}: within a hyperfine
multiplet the ratio of the sums of the intensities of all transitions from two
states with quantum numbers $F$ and $F$' are in the ratio of their statistical
weights ($2F+1$):($2F$'$+1$).

This simple theory was used in the fitting of HFS of the \ion{Co}{ii}
transitions in this work. For Co, nuclear spin $ I = 7/2 $. No perturbations
were evident.
\subsection{Experimental Details}

The spectra used in this work were obtained as part of an extensive measurement
of the \ion{Co}{ii} spectrum, by FTS with cobalt-neon and cobalt-argon hollow
cathode lamps as sources, in the wavelength region $1420-33000$ \AA\ 
($70422-3000$ cm$^{-1}$), resulting in a doubling of the number of classified
lines, over 200 new energy levels
\citep{1998ApJS..117..261P,1998PhyS...57..385P,1998PhyS...58..457P}, and new
calculations of \ion{Co}{ii} oscillator strengths \citep{1998A&AS..130..541R}.
These papers give identified line lists, including the centre-of-gravity (COG)
wavenumber and intensity of each transition, together with tables of energy
levels.  Most of the \ion{Co}{ii} lines exhibit broadening due to hyperfine
structure. The spectra used for this HFS study were acquired in the visible-UV
region with the vis-VUV Fourier transform (FT) spectrometer at Imperial College,
with resolution $0.037$~cm$^{-1}$ ($0.006$~\AA) in the visible, $0.05$~cm$^{-1}$
($0.003$~\AA) in the UV, the individual line components being limited by their
Doppler widths alone. The light source was a water-cooled hollow cathode lamp
run in either Ar at $1$~mbar  or Ne at $2-3$~mbar, with currents of
$600-700$~mA, using a pure cobalt cathode.
\subsection{Hyperfine Structure Analysis}
The determination of the $A$ and $B$ factors for the \ion{Co}{ii} energy levels
presented in this paper was carried out using a HFS analysis subroutine in the
spectrum analysis program Xgremlin \citep{Nave97}. The input information
comprises the $I$-value, the $J$-values of both levels, the known intensity
ratios of the transition components \citep{Kuhn62}, the widths of the components
(having Gaussian and Lorenzian contributions), the COG of the pattern and, where
possible (see below) an $A$ factor for one of the two levels. A large scale
study of \ion{Co}{ii} HFS is ongoing at Imperial College, and here we report a
detailed investigation of $A$ factors for the $3d^{7}(^{4}$P$)4s \;$
\Co{a}{5}{P}{}{3,2,1} levels and for three of the four $3d^{7}(^{4}$F$)4p\;$
\Co{z}{5}{D}{o}{4,3,1} levels, with immediate application in this work.

It was not possible to determine the electric quadrupole hyperfine interaction
constant $B$ to great accuracy, because the contribution of the last term in Eq.
$1$ was so small, but $B$ factors are estimated. The analysis was particularly
challenging because no previous measurements existed, but with a comprehensive
study we were able to set a starting point for the fits of line profiles in
cases where an initial guess at the $A$ factors could be made from a
characteristic line profile pattern. For example, when one of the $A$ factors 
of the levels involved in a transition is an order of magnitude greater than the
other $A$ factor, the line profile has a characteristic ``flag" pattern
\citep[examples shown in][]{1996ApJS..107..811P}. These $A$ factors could then
be used to find $A$ factors of other levels using different transitions,
building up the number of known $A$ factors.

In general, for each level studied the HFS $A$ factor was determined from the
analysis of line profiles for between $3$ and $6$ different transitions, however
in the case of the even levels reported here between $7$ and $13$ line profiles
were fitted.

The resolution of the FT spectrometer is such that the line widths are limited
solely by the Doppler widths and so, depending on the size of the $A$ factors,
the overall pattern is usually well resolved into individual components in the
IR region, but only partly so in the UV. However, even where a set of line
profiles did not appear to be resolved into individual components, the fitting
still gave consistent sets of $A$ factors.

The high number of fitted lines, the fact that the $A$ factors were relatively
large in magnitude compared to those of higher lying odd levels, and good
signal-to-noise ratio of studied lines gave reliable $A$ factors for the
\Co{a}{5}{P}{}{3,2,1} levels with uncertainties ranging between $3-7$ \%.
However, fitting of line profiles to find the $A$ factors for the four odd 
\Co{z}{5}{D}{o}{} levels was less reliable because of the smaller value of the
$A$ factor, resulting in almost no visible structure in the line profile. $A$
factors for these odd levels should be considered to be \emph{estimates}.

\begin{figure}
\resizebox{\columnwidth}{!}{\rotatebox{-90}
{\includegraphics[scale=1]{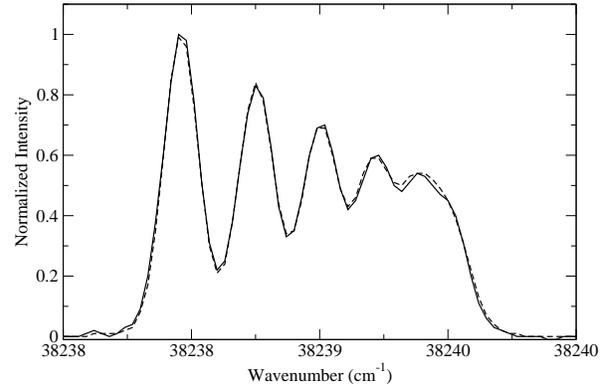}}}
\caption[]{An example of an observed (heavier trace in plot) and fitted (dashed
trace) line profile, for a \ion{Co}{ii} transition recorded by FTS: ($38238.965$
cm$^{-1}$) $2614.3530$~\AA\ transition, $3d^{7}(^{4}$P$)4s$ \Co{a}{5}{P}{}{3}
- $3d^{7}(^{4}$P$)4s$ \Co{z}{5}{D}{o}{2}.}
\label{hfs_figure}
\end{figure}

The values determined for the $A$ and $B$ splitting factors are listed in Table
\ref{hfs_table}, and they are the weighted average values from the line fits
with weighting dependent on the signal-to-noise ratio of the transitions of the
individual line fits \citep{2005MNRAS.364..705B}. The uncertainties listed in
Table \ref{hfs_table} are estimated from the range of values of each particular
$A$ factor, from the standard deviation. In Table \ref{hfs_table} the first four
columns give, for each level investigated, the configuration, term designation,
$J$-value and energy reported in \citet{1998ApJS..117..261P}. The fifth and
sixth columns give the $A$ factor and its uncertainty in mK
($1$mK=$0.001$cm$^{-1}$) determined in this work. The next column lists, where
appropriate, approximate values of $B$ factors and their uncertainty. The final
column gives the number of line profiles fitted in the process of finding that
$A$ factor. An example of a fitted line profile used in this work is shown in
Fig. \ref{hfs_figure}.\\
%
% do not forget that there must be a linebrake after rotatebox {-90} !!!
% otherwise the line disappears
\begin{table}
\renewcommand{\tabcolsep}{4.5pt}
\begin{minipage}{\linewidth}
\caption{Hyperfine structure interaction constants $A$ and $B$ for selected
energy levels of \ion{Co}{ii}. N denotes the number of profiles fitted.}
\label{hfs_table}
\begin{tabular}{@{}lrclrlrrr}
\hline
Configuration & Term &  J  & Energy & $A$ & err & $B$ & err &  N  \\
   &   &  & cm$^{-1}$  & mK & mK & mK & mK & \\
\hline
$3d^{7}(^{4}$P)$4s$ & \Co{a}{5}{P}{}{}  &  3  &  17771.506  & 40.0 & 3   & -20
& 30 & 10 \\
                    &                    &  2  &  18031.426  & 49.9 & 1.5 &  20 
& 30 & 13 \\
                    &                    &  1  &  18338.639  & 60.0 & 2   &  2
& 5  &  7 \\
  & & & & & & & & \\
$3d^{7}(^{4}$F)$4p$ & \Co{z}{5}{D}{o}{} &  4  &  46320.829  & 9.0 & 2 & 10 
& 50 &  3 \\
                    &                    &  3  &  47039.102  & 8.0  & 20  & - 
&    &  1  \\
                    &                    &  2  &  47537.362$^{1}$ &   &   &   &
& \\
                    &                    &  1  &  47848.778$^{2}$ & $<$10 & - &
-  & - &  \\ 
\hline
\end{tabular}
\medskip
Note: $^{1}$ transitions could not be reliably fitted as part of this work to
date, $^{2}$ transitions involving \Co{z}{5}{D}{o}{1} were observed to be
narrow, consistent with $A < 10$ mK, but fitting was unreliable, so an upper
limit of the $A$ factor alone is given here as a guide.
\end{minipage}
\end{table}

\section{NLTE line formation}{\label{sec:nlte}}

\subsection{The methods, model atmospheres, and stellar parameters}

The statistical equilibrium (SE) calculations for Co are carried out with the
DETAIL code \citep{Butler85}, with the radiative transfer based on the method of
accelerated lambda iteration.

Spectrum synthesis is performed using the SIU code written by J. Reetz. The
profiles of \ion{Co}{i} and \ion{Co}{ii} lines are computed with NLTE atomic
level populations from SE calculations. Radiation damping is derived from the
oscillator strengths of transitions. Damping due to quadratic Stark and van der
Waals effects follows the expressions given by \citet{1955QB461.U55......}.
However, the atomic interaction parameter $C_6$ is calculated using the line
half-widths due to elastic collisions with \ion{H}{i} computed from the theory
of \citet{1995MNRAS.276..859A}. The $\log C_6$ values and other parameteres for
each Co line investigated in this work are given in Table \ref{line_data_coi}.

The hyperfine structure of lines is taken into account by superposing
individually synthesised HFS components. Relative intensities of the components
are calculated according to the tables of \citet{1933PhRv...44..753W}.
Wavelengths are computed using the laboratory data for $A$ and $B$ factors.
HFS data for all \ion{Co}{i} levels are taken from \citet{1996ApJS..107..811P}.
For the levels of \ion{Co}{ii}, we use our new values as described in Sect.
\ref{sec:meas}. The $A$ and $B$ factors are given in the Table \ref{HFS_co}
available online.

Abundances are determined by fitting the synthetic lines to the observed line
profiles. The analysis of stars is \emph{strictly differential}\footnote{The
differential element abundance in a metal-poor star is given by $$
\mathrm{[El/H]}= \loggfestar - \loggfesun $$} relative to the Sun, i.e. any
abundance estimate derived from a single line in a spectrum of a metal-poor star
is referred to that from the corresponding solar line. This excludes the use of
oscillator strengths in calculation of stellar [Co/Fe] ratios. 

All computations, including determinations of stellar parameters, are based on
the model atmospheres computed with the MAFAGS code. These are static 1D
plane-parallel models with line blanketing computed with opacity distribution
functions of \citet{1992RMxAA..23...45K}. Convection is taken into account using
the mixing-length theory \citep{1958ZA.....46..108B} with the mixing length
parameter $\alpha = 0.5$. Some examples and tests of MAFAGS-ODF models are given
in \citet{2004A&A...420..289G}. Note that due to physical limitations of such
classical 1D LTE models, our analysis is restricted to turnoff stars. Only one
giant, HD 122563, is included for purposes of comparison with other studies.

Basic parameters for the program stars were adopted from
\citet{2004AN....325....3F}, \citet{2004A&A...413.1045G,2006A&A...451.1065G},
and \citet{2008A&A...478..529M}. These analyses make use of the following
methods. The effective temperatures were derived by fitting the observed
H$_\alpha$ and H$_\beta$ profiles under LTE assumption. The surface gravities
were calculated using parallaxes $\pi$ measured by \emph{Hipparchos}
\citep{1997yCat.1239....0E} and bolometric corrections were taken $BC$ from
\citet{1995A&A...297..197A}. Masses were determined from the tracks of
\citet{2000ApJ...532..430V} by interpolating in the $M_{\rm bol}$ - $\Teff$
diagram. The iron abundances and microturbulent velocities were obtained from
\ion{Fe}{ii} line profile fitting in LTE. For the thick disk and halo stars, the
abundances of $\alpha$-elements are enhanced by $\sim 0.4$ dex. The population
identification is based on stellar kinematic properties, ages, and abundance
ratios [Al/Mg]. All stellar parameters are listed in Table \ref{stel_param_gen}.

\begin{table}
\begin{small}
\renewcommand{\footnoterule}{}  
\renewcommand{\tabcolsep}{2.5pt}
\caption{Lines of \ion{Co}{i} and \ion{Co}{ii} selected for solar and stellar
abundance calculations. $N$ denotes a number of HFS components for a line, and
the multiplet is specified in the 2-d column. $\Elow$ is the excitation energy
of the lower level of a transition. $W_\lambda$ refers to the line equivalent
width. An asterisk in the wavelength entry refers to the lines, which are not
used in determination of the solar Co abundance due to blending or inaccurate
$\log gf$'s.}
\label{line_data_coi}
\begin{center}
\begin{tabular}{ll|rrlcc|rrlllll}
\hline 
 ~~~~~$\lambda$ & Mult. & $N$ & $\Elow$ & Lower & Upper & $W_\lambda$ &
\multicolumn{2}{c}{$\log gf^a$} & & $\log C_6$ \\
 ~~~~[\AA] &  &  & [eV] & level & level & [\mA] & \multicolumn{2}{c}{} & & \\
\hline
%\hline \multicolumn{13}{l}
\ion{Co}{i} & & & & & & & \multicolumn{2}{c}{} & \\
3845.470* &  34 & 4 &  0.92  &  \Co{a}{2}{F}{ }{7/2} & \Co{y}{2}{G}{o}{9/2}
& 120     &   0.01  & 2 & & --31.5 \\
3957.930* &  18 & 3 &  0.58  &  \Co{b}{4}{F}{ }{5/2} & \Co{z}{4}{D}{o}{5/2}
& 60      & --2.07  & 2 & & --31.7 \\
4020.905 &  16 & 6 &  0.43  &  \Co{b}{4}{F}{ }{9/2} & \Co{z}{4}{F}{o}{9/2}
& 76      & --2.04  & 1 & & --31.8  \\
4066.360* &  30 & 3 &  0.92  &  \Co{a}{2}{F}{ }{7/2} & \Co{y}{4}{D}{o}{7/2}
& 59      & --1.60  & 1 & & --31.6  \\
4110.530 &  29 & 6 &  1.05  &  \Co{a}{2}{F}{ }{5/2} & \Co{z}{2}{F}{o}{5/2}
& 95      & --1.08  & 2 & & --31.6  \\
4121.320 &  28 & 4 &  0.92  &  \Co{a}{2}{F}{ }{7/2} & \Co{z}{2}{G}{o}{9/2}
& 131     & --0.30  & 1 & & --31.7  \\
4792.862* & 158 & 6 &  3.25  &  \Co{z}{6}{G}{o}{7/2} & \Mn{e}{6}{F}{ }{5/2}
& 34      & --0.07  & 3 & & --30.5  \\
4813.480* & 158 & 7 &  3.22  &  \Co{z}{6}{G}{o}{9/2} & \Mn{e}{6}{F}{ }{7/2}
& 44      & ~~0.05  & 3 & & --30.5  \\
4867.870* & 158 &10 &  3.17  &  \Co{z}{6}{G}{o}{11/2} & \Mn{e}{6}{F}{}{9/2}
& 60      & ~~0.23  & 3 & & --30.6  \\
5212.691 & 170 & 9 &  3.51  &  \Co{z}{4}{F}{o}{9/2} & \Co{f}{4}{F}{ }{9/2}
& 25      & --0.11  & 2 & & --30.5  \\
5280.631 & 172 & 8 &  3.63  &  \Co{z}{4}{G}{o}{9/2} & \Co{f}{4}{F}{ }{7/2}
& 20      & --0.03  & 2 & & --30.4  \\
5301.047 &  39 & 5 &  1.71  &  \Co{a}{4}{P}{ }{5/2} & \Co{y}{4}{D}{o}{5/2}
& 21      & --1.94  & 1 & & --31.5  \\
5331.460 &  39 & 5 &  1.78  &  \Co{a}{4}{P}{ }{1/2} & \Co{y}{4}{D}{o}{3/2}
& 17      & --1.99  & 1 & & --31.5  \\
5352.049 & 172 & 9 &  3.58  &  \Co{z}{4}{G}{o}{11/2} & \Co{f}{4}{F}{}{9/2}
& 26      &   0.06  & 2 & & --30.5  \\
5369.590* &  39 & 4 &  1.74  &  \Co{a}{4}{P}{ }{3/2} & \Co{y}{4}{D}{o}{5/2}
& 44      & --1.59  & 1 & & --31.5  \\
5483.340 &  39 & 7 &  1.71  &  \Co{a}{4}{P}{ }{5/2} & \Co{y}{4}{D}{o}{7/2}
& 51      & --1.41  & 1 & & --31.5  \\
5647.234 & 112 & 6 &  2.28  &  \Co{a}{2}{P}{ }{3/2} & \Co{y}{2}{D}{o}{5/2}
& 14      & --1.56  & 2 & & --31.4  \\
6189.000* &  37 & 9 &  1.71  &  \Co{a}{4}{P}{ }{5/2} & \Co{z}{4}{D}{o}{5/2}
& 11      & --2.45  & 2 & & --31.7 \\
6454.990 & 174 & 8 &  3.13   &  \Co{z}{4}{D}{o}{7/2} & \Co{e}{6}{F}{ }{9/2}
& 15      & --0.25  & 2 & & --30.4  \\
6814.950* &  54 & 5 &  1.96  &  \Co{b}{4}{P}{ }{3/2} & \Co{z}{4}{D}{o}{3/2}
& 19      & --1.90  & 2 & & --31.7 \\
7417.380* &  89 & 8 &  2.04  &  \Co{a}{2}{D}{ }{3/2} & \Co{z}{4}{D}{o}{5/2}
& 11      & --2.07  & 2 & & --31.7  \\
7712.661* & 126 & 8 &  2.04  &  \Co{b}{2}{P}{ }{3/2} & \Co{z}{2}{D}{o}{5/2}
& 10      & --1.57  & 2 & & --31.5  \\
\ion{Co}{ii} & & & & & & & \multicolumn{2}{c}{} & & & \\
3501.730  &  2 & 8 &  2.19  &  \Co{a}{5}{P}{}{3} & \Mn{z}{5}{D}{o}{4}
& 86      & --1.22  & 4 & & --32.2  \\ 
\noalign{\smallskip}\hline\noalign{\smallskip}
\end{tabular}
\end{center}
$^a$ References: ~~~(1) \citet{1999ApJS..122..557N}; (2)
\citet{1982ApJ...260..395C}; (3) \citet{1996yCat.6010....0K}; (4)
\citet{1998A&AS..130..541R}
\end{small}
\end{table}
\begin{table*}
\begin{centering}
\begin{minipage}{\textwidth}
\renewcommand{\footnoterule}{} 
\tabcolsep1.5mm \small \caption{Stellar parameters and their estimated errors
for the selected sample. In the \emph{Ref} column, the sources of adopted
stellar parameters are given.} 
\vspace{5mm}
\label{stel_param_gen}
\begin{tabular}{lrrr@{$\,\pm\,$}lr@{$\,\pm\,$}lcrrrlrrrrlcc}
\hline\noalign{\smallskip}
Object & HIP & Instr. & \multicolumn{2}{c}{$\Teff$} & \multicolumn{2}{c}{$\log
g$} & $\xi_{\rm t}$ & [Fe/H] & [Mg/Fe] & $\pi/\sigma_\pi$ & Population &
Ref.\footnote{References:~~(1) \citet{2006A&A...451.1065G}; (2)
\citet{2004A&A...413.1045G}; (3) \citet{2008A&A...478..529M}; (4)
\citet{2004AN....325....3F}; (5) \citet{2003A&A...407..691K}} & $N_{\rm{Co}}$ &
\multicolumn{2}{c}{[Co/Fe]} \\
  &  &  & \multicolumn{2}{c}{[K]} & \multicolumn{2}{c}{ } & [km/s] &   &   &   
&  &  & & NLTE & LTE \\
\noalign{\smallskip}\hline\noalign{\smallskip}
HD 19445         &  14594 & FOCES & 5985 &  80 & 4.39 & 0.05 & 1.5 & $-1.96$ &
0.38 & 22.7 & Halo & 1  & 1 & $0.64$ & $0.08$ \\
HD 25329         &  18915 & FOCES & 4800 &  80 & 4.66 & 0.08 & 0.6 & $-1.84$ &
0.42 & 50.1 & Thick Disk? & 4 & 1 & $-0.02$ & $-0.72$ \\
HD 29907         &  21609 & UVES & 5573 & 100 & 4.59 & 0.09 & 0.9 & $-1.60$ &
0.43 & 17.3 & Halo? & 2   & 2 & $0.66 \pm 0.06$ & $0.19 \pm 0.05$ \\
HD 34328         &  24316 & UVES & 5955 &  70 & 4.47 & 0.07 & 1.3 & $-1.66$ &
0.42 & 14.4 & Halo  & 2   & 3 & $0.67 \pm 0.01$ & $0.11 \pm 0.16$ \\
HD 61421        &  37279  & UVES & 6510 & 100 & 3.96 & 0.05 & 1.8 & $-0.03$ &
0.0 & 324.9 & Thin Disk & 5 & 8 & $-0.09 \pm 0.07$ & $-0.15 \pm 0.1$ \\
HD 84937         &  48152 & UVES & 6346 & 100 & 4.00 & 0.08 & 1.8 & $-2.16$ &
0.32 & 11.7 & Halo  &  1  & 3 & $0.66 \pm 0.03$ & $0.18 \pm 0.01$ \\
HD 102200        &  57360 & UVES & 6120 &  90 & 4.17 & 0.09 & 1.4 & $-1.28$ &
0.34 & 10.5 & Halo  &  2  & 3 & $0.51 \pm 0.06$ & $0.03 \pm 0.11$ \\
HD 103095        &  57939 & FOCES & 5110 & 100 & 4.69 & 0.10 & 1.0 & $-1.35$ &
0.26 & 140.0 & Halo  &  1 & 1 & $-0.01$ & $-0.11$  \\
BD$-4^\circ3208$ &  59109 & UVES & 6310 &  60 & 3.98 & 0.21 & 1.5 & $-2.23$ &
0.34 & 3.7 & Halo    &  2 & 2 & $0.71 \pm 0.08$ & $0.13 \pm 0.01$ \\
HD 122196        &  68464 & UVES & 5957 &  80 & 3.84 & 0.11 & 1.7 & $-1.78$ &
0.24 & 7.4 & Halo    &  2 & 2 & $0.52 \pm 0.02$ & $-0.12 \pm 0.06$ \\
HD 122563        &  68594 & UVES & 4600 & 120 & 1.50 & 0.20 & 1.9 & $-2.51$ &
0.45 & 5.2 & Halo    &  3 & 3 & $0.57 \pm 0.02$ & $-0.07 \pm 0.06$ \\
HD 134169        &  74079 & FOCES & 5930 & 200 & 3.98 & 0.1 & 1.8 & $-0.86$ &
0.53 & 15.1 & Thick disk & 1 & 3 & $0.45 \pm 0.02$ & $0.18 \pm 0.14$ \\
HD 148816        &  80837 & FOCES & 5880 & 200 & 4.07 & 0.07 &  1.2 & $-0.78$ &
0.36 & 27 & Thick disk & 2 & 6 & $0.38 \pm 0.03 $ & $0.19 \pm 0.06$ \\
HD 184448        &  96077 & FOCES & 5765 & 200 & 4.16 & 0.07 &  1.2 & $-0.43$ &
0.47 & 30.4 & Thick disk & 2 & 11 & $0.34 \pm 0.04$ & $0.27 \pm 0.03$ \\
HD 140283        &  76976 & UVES & 5773 &  60 & 3.66 & 0.05 & 1.5 & $-2.38$ &
0.43 & 18.0 & Halo       & 2 & 3 & $0.85 \pm 0.02$ & $0.14 \pm 0.01$ \\
G 20-8           &  86443 & FOCES & 6115 &  80 & 4.20 & 0.20 & 1.5 & $-2.19$ &
0.45 & 5.1   & Halo       & 1 & 2 & $0.76 \pm 0.06$ & $0.28 \pm 0.15$ \\
G 64-12           & 66673 & UVES  & 6407 &  80 & 4.20 & 0.20 & 2.3 & $-3.12$ &
0.33 & 1.3   & Halo  & 1 & 1 & $0.8 \pm 0.1 $ & $0.19 \pm 0.1$ \\
HD 200580        & 103987 & FOCES & 5940 &  80 & 3.96 & 0.06 & 1.4 & $-0.82$ &
0.46 & 13.8 & Thick Disk & 2 & 2 & $0.38 \pm 0.04$ & $0.12 \pm 0.1$ \\
\noalign{\smallskip}\hline\noalign{\smallskip}
\end{tabular}
\end{minipage}
\end{centering}
\end{table*}

\subsection{Atomic model}\label{sec:atmmod}

The model of the Co atom is constructed as follows. The number of levels and
transitions is $246$ and $6027$ for \ion{Co}{i}, and $165$ and $2539$ for
\ion{Co}{ii}. The energy separation of the highest excited level from the
continuum is $0.4$ eV for \ion{Co}{i} and $\sim 3$ eV for \ion{Co}{ii}.
Energies of levels and wavelengths of transitions are taken from
\citet{1996ApJS..107..761P} and \citet{1998ApJS..117..261P}. The oscillator
strengths are adopted from the Kurucz' database \citep{1995KurCD..23.....K},
which includes laboratory measurements and data calculated using scaled
Thomas-Fermi-Dirac radial wavefunctions. A complete Grotrian diagram for Co is
available online.

Hyperfine splitting of the levels was not included in the SE calculations,
since \emph{relative} populations of HFS components are thermal. Moreover, for
the uppermost terms above $6.3$ eV in \ion{Co}{i} the fine structure is not
maintained. These terms are represented by a single level with a weighted mean
of statistical weights and ionization frequencies of their fine structure
levels. The transitions between two combined levels are also grouped, and the
oscillator strength of a resulting line is the average of $\log gf$'s weighted 
according to the appropriate lower-level statistical weights.

Since no quantum-mechanical calculations of photoionization for \ion{Co}{i}
and \ion{Co}{ii} are available, hydrogenic approximation is used. The
cross-sections are computed from the formula of Kramer 
\citep{1935MNRAS..96...77M} corrected for the ion charge \citep[see, e.g.,
][]{2003ASPC..288...99R} and using the \emph{effective} hydrogen-like main
quantum number. For cross-sections of collisional ionization, we use the formula
of \citet{1962amp..conf..375S}. The rates of allowed and forbidden bound-bound
transitions due to collisions with electrons are calculated using the formulae
of \citet{1962ApJ...136..906V} and \citet{1973asqu.book.....A}, respectively.

The rates of bound-bound and bound-free transitions due to inelastic
collisions with \ion{H}{i} atoms are computed from the formula of
\citet{1969ZPhy..225..483D} in the version of \citet{1984A&A...130..319S} and
are multiplied by a scaling factor $\SH = 0.05$.
This value gave the smallest fitted-abundance spread in our previous work on SE
of Mn in stellar atmospheres \citep{2007A&A...473..291B,2008A&A...492..823B}.
Most abundance analyses favour similar small values of $\SH$
\citep{1996ASPC..108..140M,2004MNRAS.350.1127A,2008A&A...486..985S,
2008A&A...486..303S}. To the best of our knowledge, there are only a few
indications in the literature that $\SH$ is \emph{larger} than unity
\citep{1999A&A...350..955G,2003A&A...407..691K}. Unfortunately, these cases
refer to the most prominent member of the Fe-group, iron.

In subsequent sections, we study the influence of photoionization cross-sections
and inelastic collisions with hydrogen on populations of atomic levels in Co.
However, it should be kept in mind that values for cross-sections have only
order-of-magnitude accuracy.

\section{Results}

\subsection{Statistical equilibrium of Co}

\begin{figure*}
\vspace{-4mm}
\hbox{\resizebox{60mm}{!}{\includegraphics[trim=1.2cm 0.4cm 0.7cm
1cm, clip]{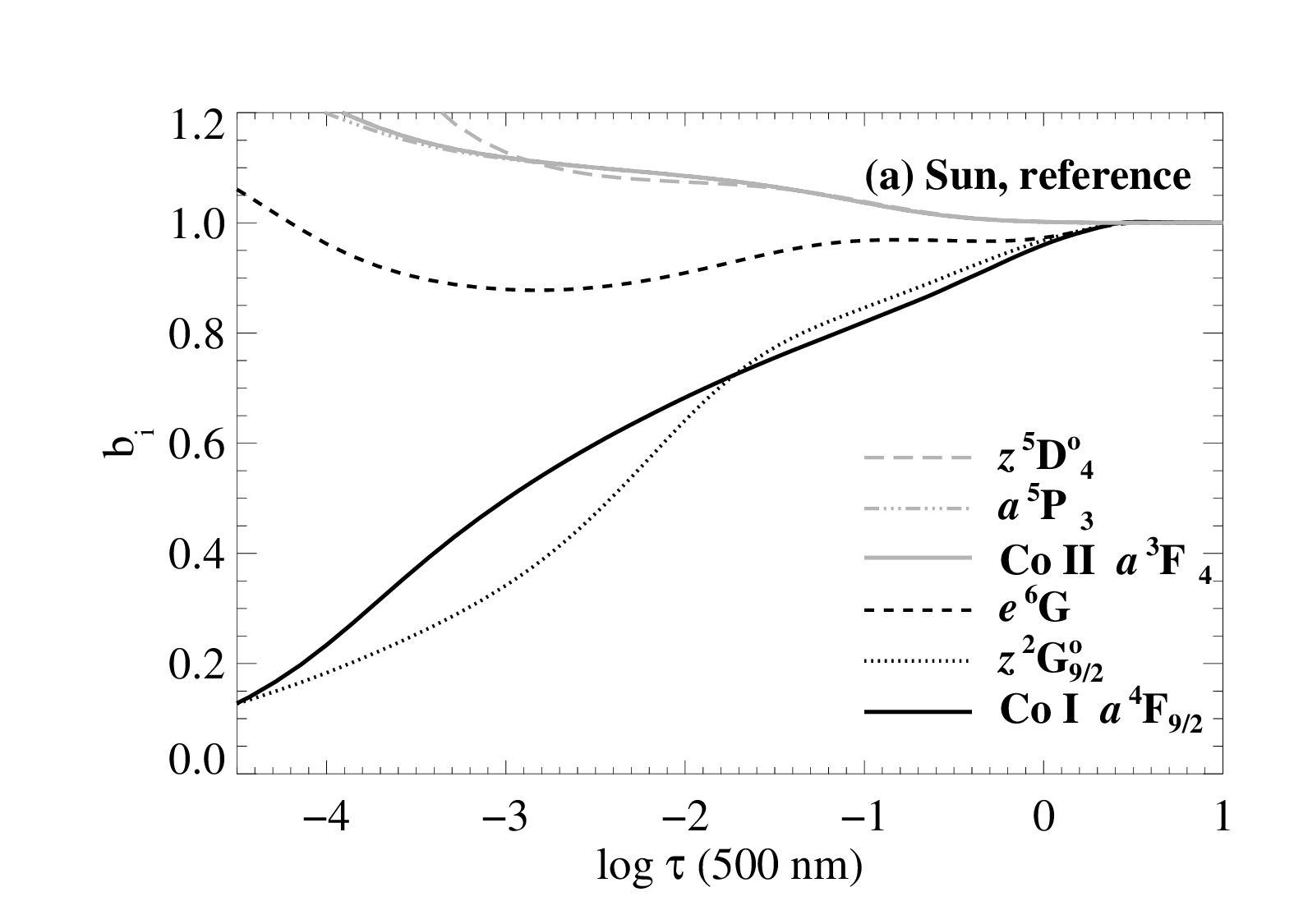}}\hfill
      \resizebox{60mm}{!}{\includegraphics[trim=1.2cm 0.4cm 0.7cm
1cm, clip]{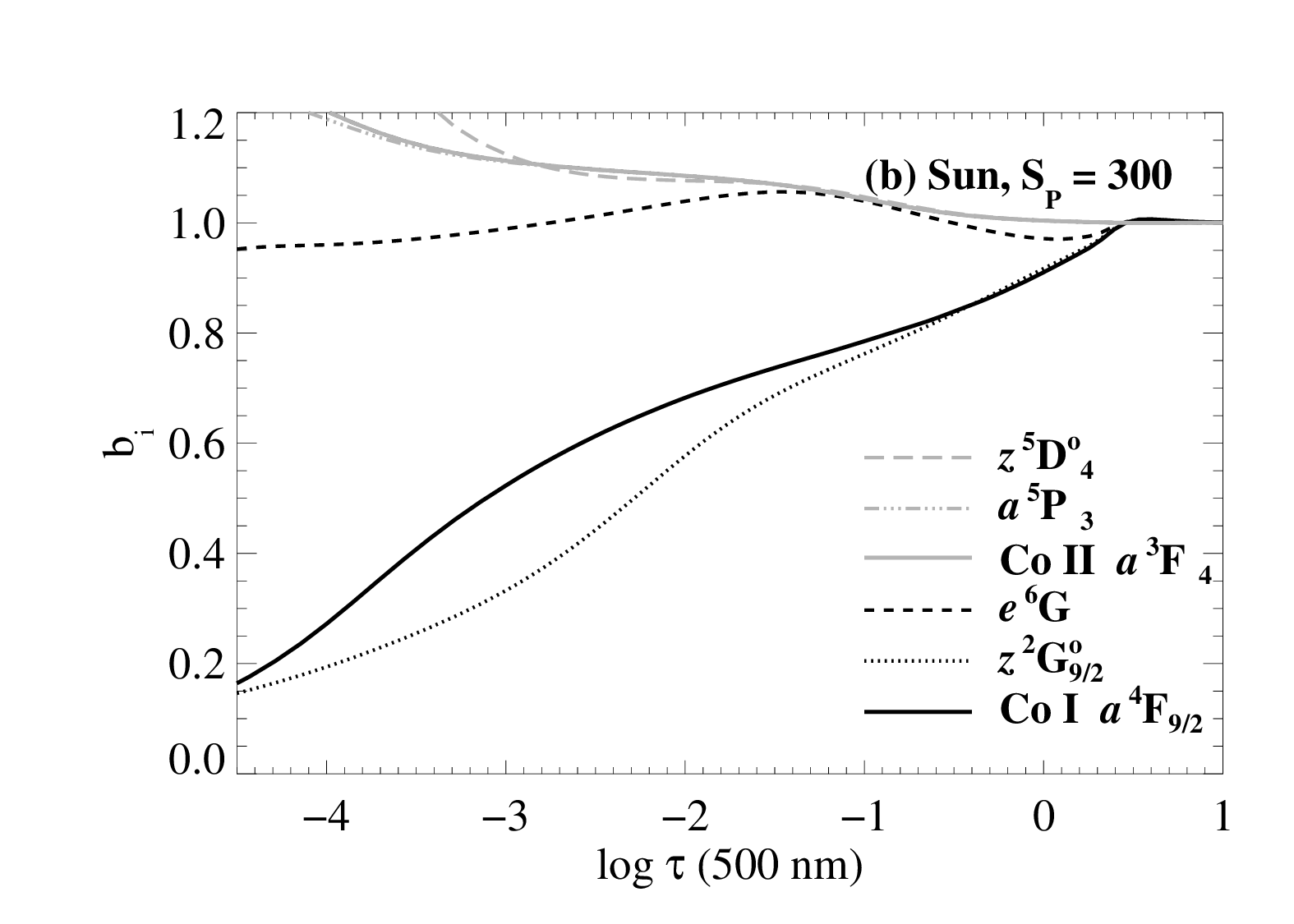}}\hfill
      \resizebox{60mm}{!}{\includegraphics[trim=1.2cm 0.4cm 0.7cm
1cm, clip]{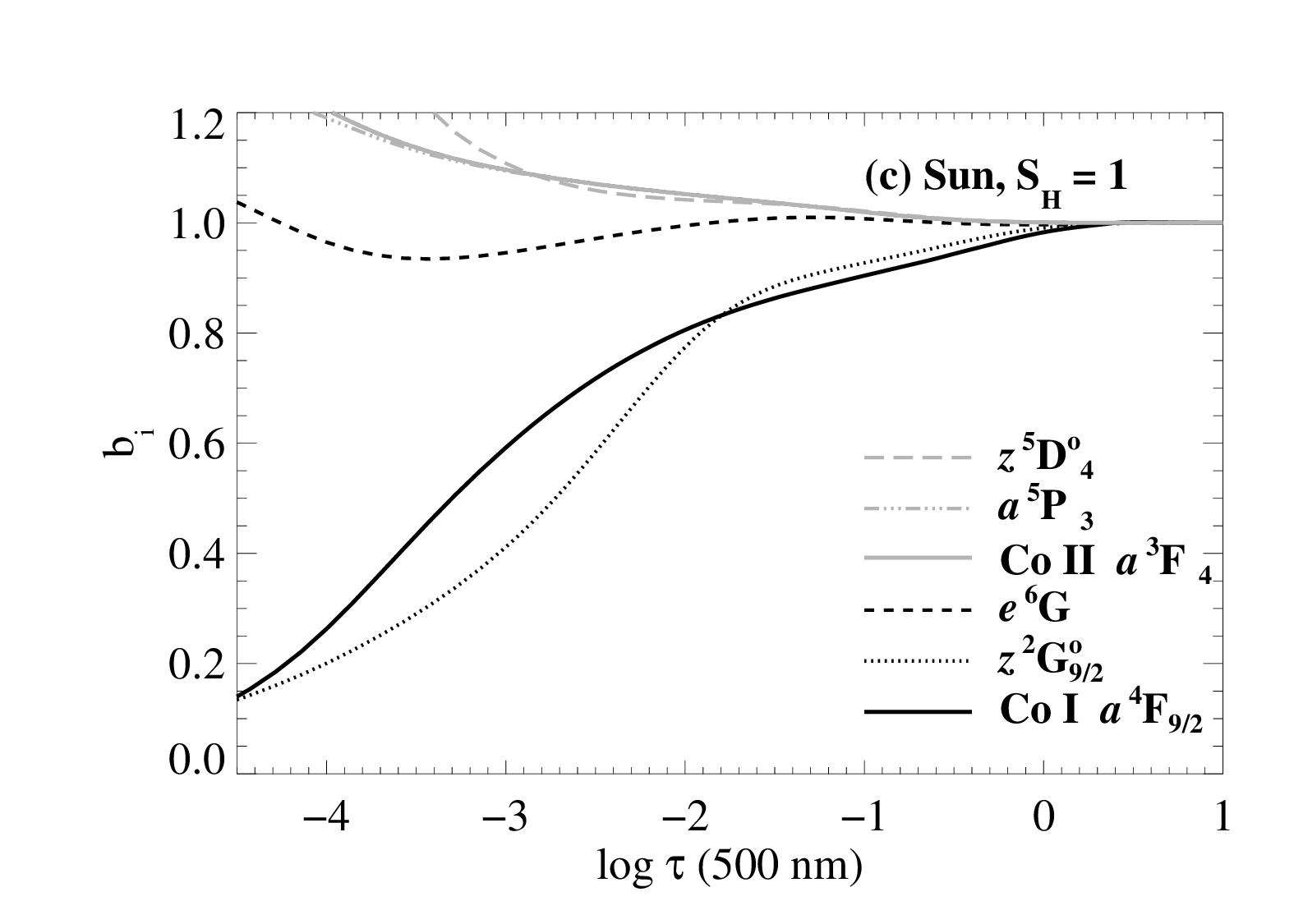}}}
%\vspace{-4mm}
\hbox{\resizebox{60mm}{!}{\includegraphics[trim=1.2cm 0.4cm 0.7cm
1cm, clip]{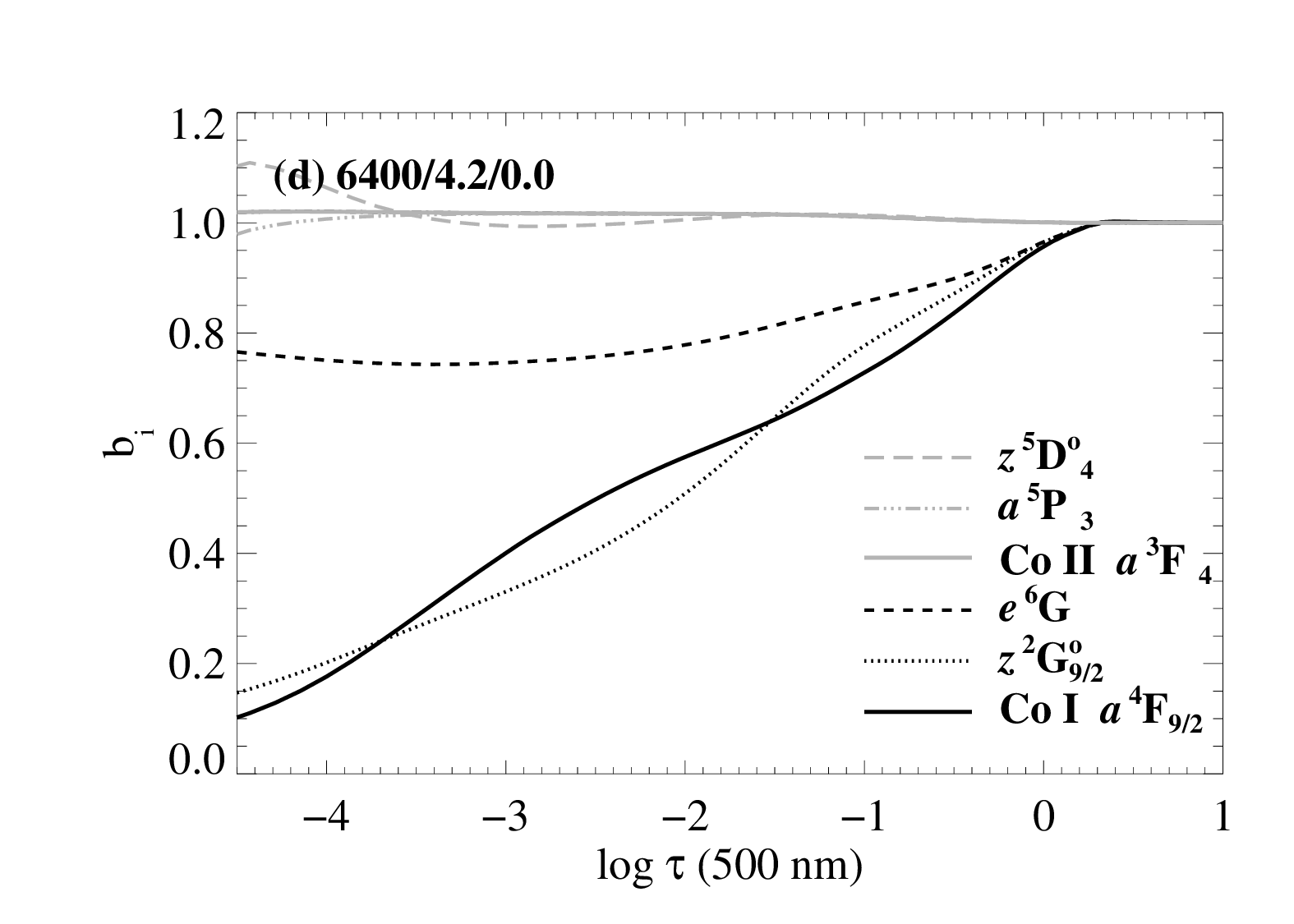}}\hfill
      \resizebox{60mm}{!}{\includegraphics[trim=1.2cm 0.4cm 0.7cm
1cm, clip]{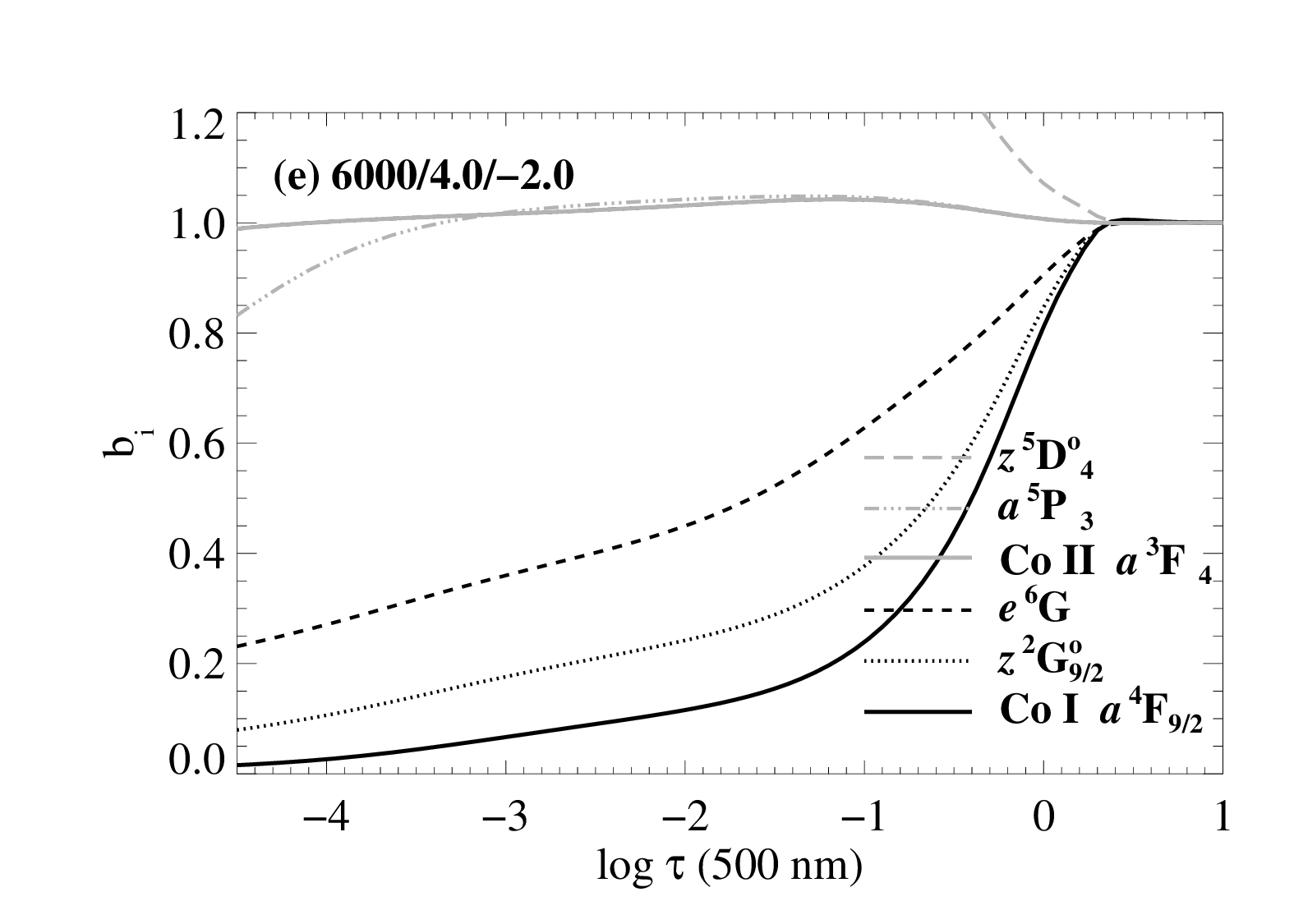}}\hfill
      \resizebox{60mm}{!}{\includegraphics[trim=1.2cm 0.4cm 0.7cm
1cm, clip]{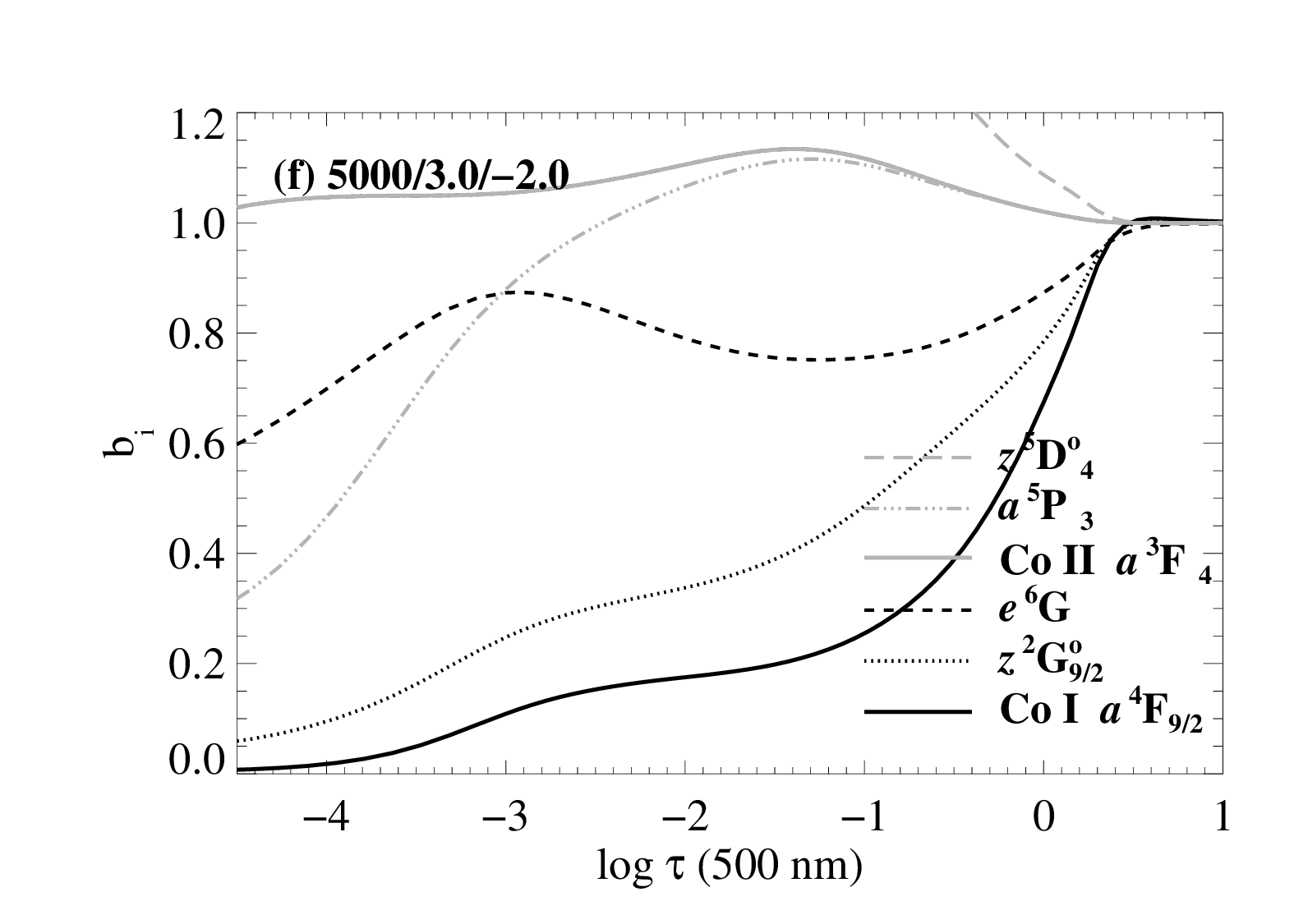}}}
\caption{Departure coefficients $b_i$ of selected \ion{Co}{i} and \ion{Co}{ii}
levels for the solar model and for several stellar models from the grid.}
\label{bfac_co}
\end{figure*}

To discuss the excitation-ionization equilibrium of Co in stellar atmospheres,
we use the notation of the \emph{departure coefficient} $b_i$, which is defined
as $ b_i = N^{\rm NLTE}_i/N^{\rm LTE}_i $ with $N_i$ the number density of
atoms in the level $i$. The $b_i$-factors as a function of optical depth at
$500$ nm are presented in Fig. \ref{bfac_co} for the solar model atmosphere and
several stellar models from the grid. A few selected levels, typical for their
depth dependence, are indicated by heavier black traces; the remaining bulk of
the \ion{Co}{i} and \ion{Co}{ii} levels are indicated by grey lines.

\subsubsection{The Sun}

Fig. \ref{bfac_co}a shows the departure coefficients, calculated with our
\emph{reference} model of Co for the Sun. The overall underpopulation of all
\ion{Co}{i} levels is due to the \emph{overionization} mainly from the levels of
$2 - 3$ eV excitation, e.g.\ \Co{z}{2}{G}{o}{} with threshold wavelength
$3151$\AA. The \ion{Co}{i} ground state is not subject to strong overionization.
But its separation from the first metastable level is only $0.4$ eV, which
favours a very efficient collisional interaction. Strong collisional and
radiative coupling also persists between other \ion{Co}{i} levels, because the
total number of transitions in \ion{Co}{i} is very large. Thus, the distribution
of $b$-factors up to $\opd \sim -2$ has a very regular structure, with levels of
ever decreasing energy gap from the continuum showing smaller deviations of
$b_i$ from unity.

In the higher layers, spontaneous transitions in strong lines of \ion{Co}{i} 
slightly modify this simple pattern. At $\opd \leq -1.5$, the medium becomes 
transparent to the radiation in transitions between the low metastable levels 
and the levels of odd doublet and quartet term systems with $\sim 4$ eV
excitation energy, that leads to a depletion of the upper levels. One such
transition, \Co{a}{2}{F}{}{7/2} $\rightarrow$ \Co{z}{2}{G}{o}{9/2}, is of
particular interest, because the corresponding spectral line $4121$ \AA\ is
the only observable line of \ion{Co}{i} in our spectra of very metal-poor stars.

A large number of strong transitions with UV wavelengths in \ion{Co}{i} connect
the low levels with $0 - 3$ eV and the levels with $5 - 6$ eV excitation
energies $E_i$. Hence, at depths with $-1 < \opd < 0.2$ there is also
\emph{overexcitation} of the upper levels. This maintains nearly constant $b_i$
of upper levels at $-1 < \opd < 0$, in spite of increasing overionization (e.g.
level \Co{e}{6}{G}{}{} in Fig. \ref{bfac_co}a). This process barely affects
populations of the lower levels due to the small Boltzmann factor, and dominance
of photoionization from these levels.

The majority of the uppermost levels with $E_i > 6.5$ eV are underpopulated
relative to the continuum. They interact much stronger radiatively with the low
levels and with each other by means of collisions than with the \ion{Co}{ii}
ground state, \Co{a}{3}{F}{}{4}. Only for a few highest levels, with threshold
ionization energies similar to the mean kinetic energy of the electrons in the
solar atmosphere, does the collisional coupling to the \Co{a}{3}{F}{}{4} level
prevail. Thus, they are in thermal equilibrium with \Co{a}{3}{F}{}{4}. 

The ground state of \ion{Co}{ii} is also affected by NLTE, although departures
in the line formation region are not large. Low-excitation levels of
\ion{Co}{ii} are in thermal equilibrium with its ground state. But, most of the 
higher \ion{Co}{ii} levels are overpopulated due to non-equilibrium excitation
processes. This NLTE mechanism is typical for \emph{majority} ions, i.e. ions of
the dominant ionization stage of an element.

The tests performed with a smaller model constructed with $246$ \ion{Co}{i}
levels and closed by the \ion{Co}{ii} ground state indicate poor performance of
such a \emph{reduced} atomic model. This changes ionization and excitation
equilibria in both ions, which is manifested in spurious overpopulation of the
\ion{Co}{ii} ground state and stronger thermalization of \ion{Co}{i} levels.
Thus, reduction of the atomic model, as is usually done for 3D NLTE
calculations, must be performed with extreme caution.

The uncertain parameters in our models are cross-sections for photoionization
and collisions with neutral hydrogen atoms. Fig. \ref{bfac_co}b demonstrates the
effect of stronger photoionization on the atomic level populations; here,
cross-sections are multipled by a scaling factor $\SP = 300$. The main
difference with the reference model (Fig. \ref{bfac_co}a) is that depopulation
of all levels is amplified at $-1 < \opd < 0.4$. The influence of collisions
with \ion{H}{i} is fully understood from Fig. \ref{bfac_co}c. When the standard
Drawins formula is used with the scaling factor $\SH = 1$, all levels at line
formation depths $-2 > \opd > 0$ are more strongly coupled to the continuum, but
this scaling factor is too small for LTE to be reached completely.

\subsubsection{Metal-poor stars}

The character of processes leading to NLTE effects in Co does not change for
stellar atmospheres with parameters of cool turnoff stars, which form the basis
of our abundance analysis.

Fig. \ref{bfac_co}d demonstrates the effect of \emph{increasing temperature} on
the departure coefficients of selected \ion{Co}{i} and \ion{Co}{ii} levels. At
$\Teff = 6400$ K, the ionization equilibrium \ion{Co}{i}/\ion{Co}{ii} is
somewhat different from that of the solar model. Nearly $98$\% of Co atoms are
in a singly ionized stage, hence the ground state of \ion{Co}{ii} is not
sensitive to overionization from the \ion{Co}{i} levels. The level
$\Co{a}{3}{F}{}{4}$ keeps its thermodynamic equilibrium value. However,
departure coefficients in \ion{Co}{i} show larger deviations from unity. The
reason is that a stellar flux maximum is shifted to shorter wavelengths, thus
overionization affects even the lowest metastable levels with $E_i \sim 1$ eV
and $\lambda_{\rm thr} \sim 1800$ \AA. In contrast, in the solar model
overionization is important only for the levels with $E_i \sim 3$ eV.

The influence of photoionization is more pronounced with \emph{decreasing
metallicity} (Fig. \ref{bfac_co}e). The dramatic depopulation of all levels
occurs at the depths of line formation, $-1 < \opd < 0$. The overionization from
the lowest levels is balanced by increased net recombination to the higher
levels that results in a small overpopulation of the latter at $\opd \sim 0$. In
the higher layers, $\opd \sim -1$ even an overpopulation of the \ion{Co}{ii}
ground state develops. Also, we note an increasing importance of line pumping in
\ion{Co}{ii}. For example, the level $\Co{z}{5}{D}{o}{4}$ has $b_i \gg 1$
already at $\opd \sim -0.3$. This is characteristic of all \ion{Co}{ii} levels
with $E_i > 5$ eV. Hence, we expect \emph{substantial NLTE effects in the lines
of both \ion{Co}{i} and \ion{Co}{ii}} in metal-poor stars.
\begin{figure}
\begin{center}
\resizebox{0.9\columnwidth}{!}{\includegraphics[scale=1]{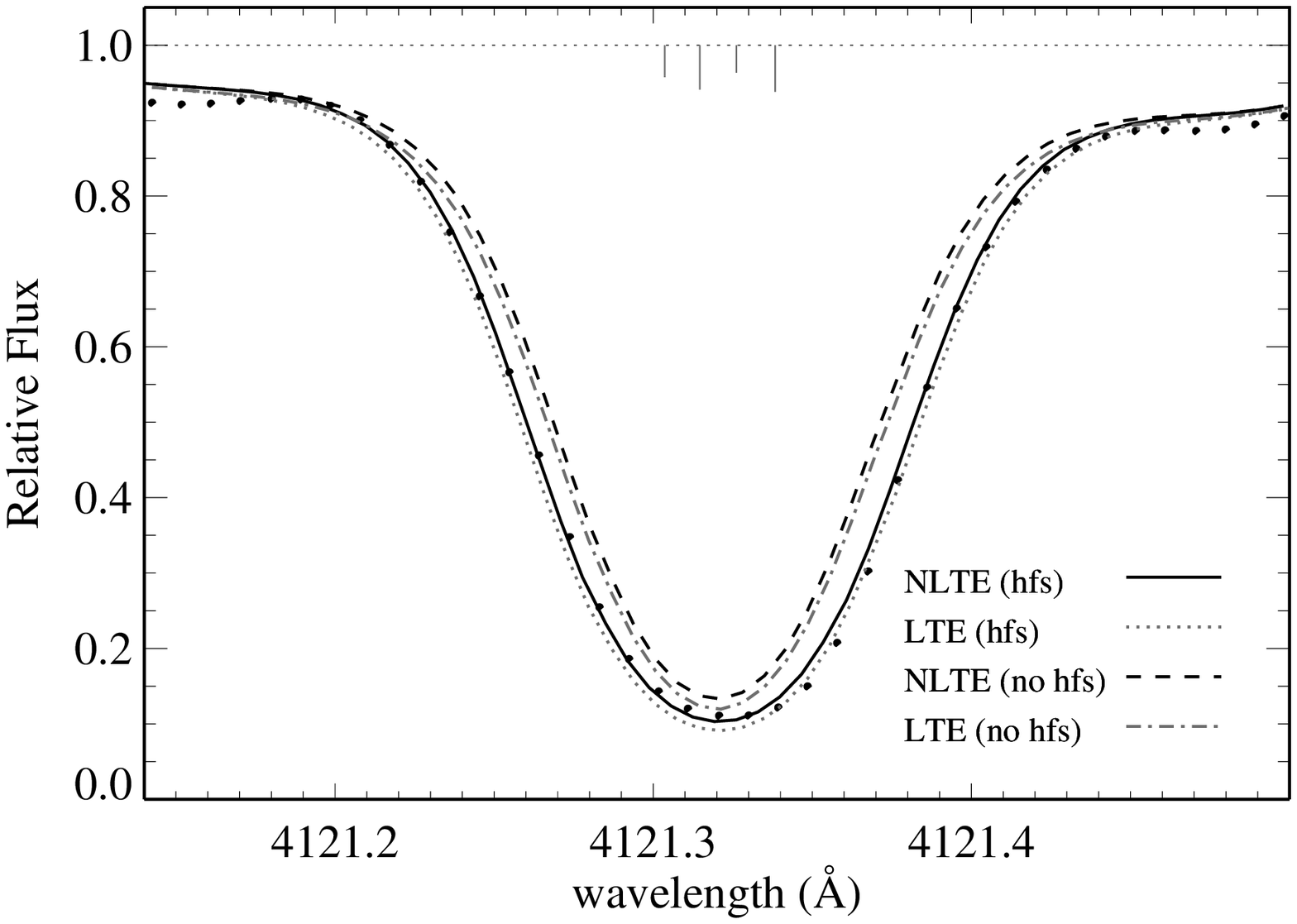}}
\hfill
\resizebox{0.9\columnwidth}{!}{\includegraphics[scale=1]{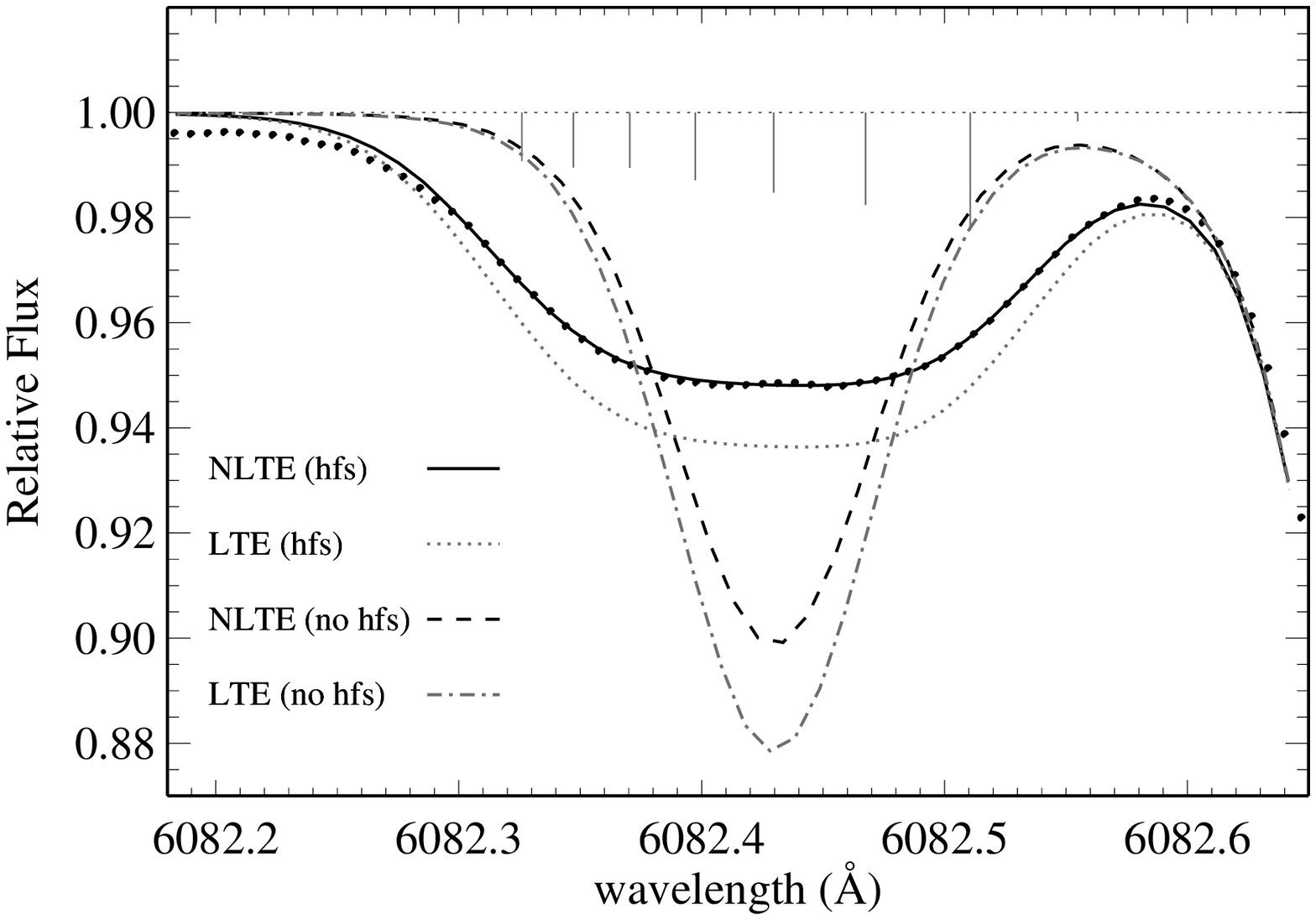}}
\hfill
\resizebox{0.9\columnwidth}{!}{\includegraphics[scale=1]{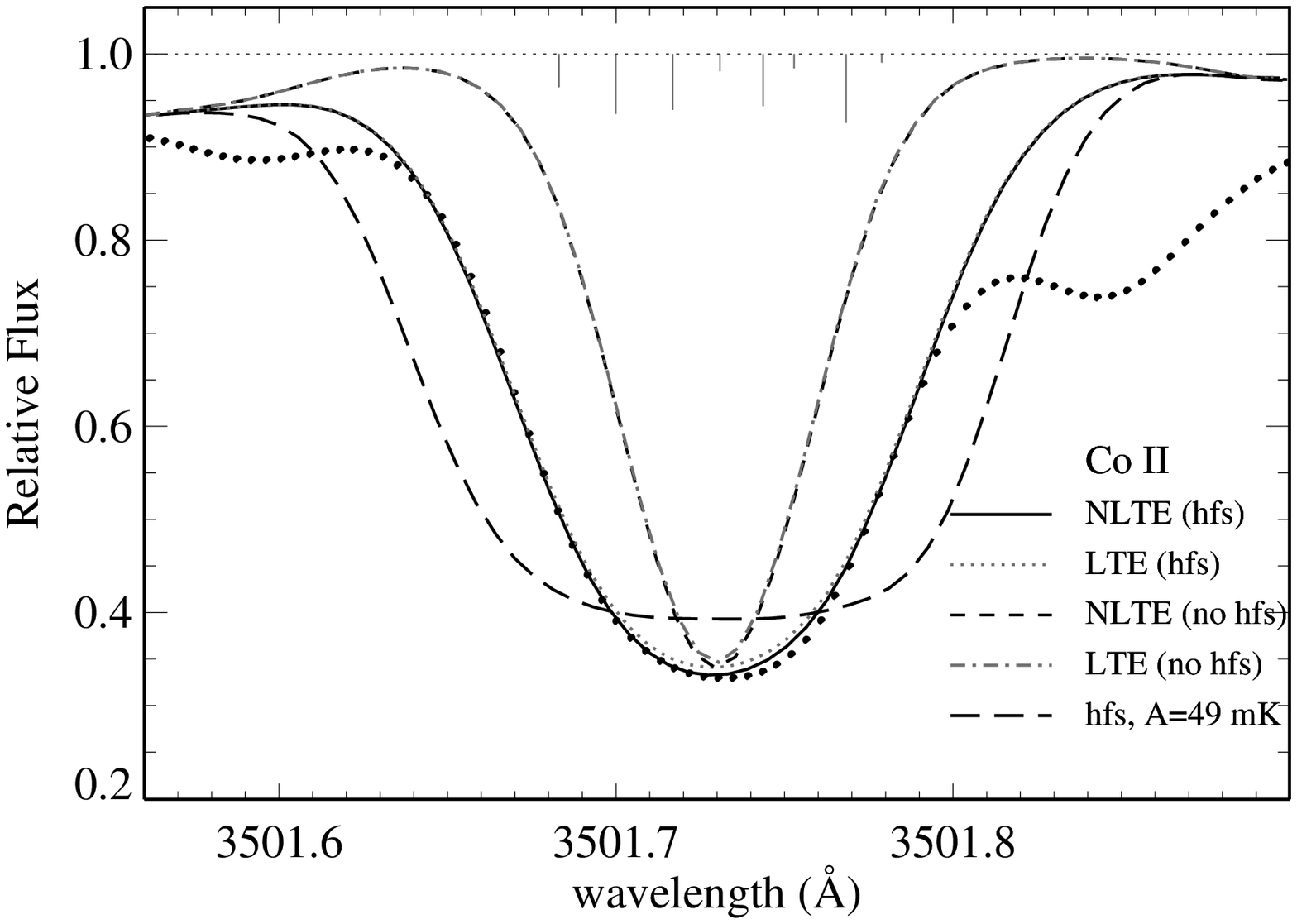}}
\caption{The lines of \ion{Co}{i} (4121, 6082 \AA) and \ion{Co}{ii} (3501
\AA) in the solar flux spectrum (filled circles). Synthetic NLTE and LTE
profiles with and without HFS are labeled correspondingly. Wavelength positions
and relative line strengths of the HFS components are indicated. Note that the
relative HFS component intensities are not on the same scale in figures. The
long-dashed trace on the bottom figure indicates the NLTE profile computed with
$A = 49$ mK for the lower level \Co{a}{5}{P}{}{3} (our reference value is $A =
40$ mK).} 
\label{ltenlte_co}
\end{center}
\end{figure}

A combination of low effective temperature, low gravity, and low metal content
leads to a rather disorderly behavior of departure coefficients for different
levels (Fig. \ref{bfac_co}f). Such stellar parameters are representative of
giants where the collisional interaction is much weaker compared to dwarfs. Each
level of \ion{Co}{i} is no longer coupled to the bulk of levels with similar
energies, but displays a distinct behaviour that is determined by the
competition of radiative b-f and b-b processes. As a result, \emph{NLTE line
formation for giants and dwarfs can be different for transitions involving
different levels}. This is evident by comparing the runs of $b_i$ for the levels
of \Co{e}{6}{G}{}{}, \Co{z}{2}{G}{o}{9/2}, and \Co{a}{3}{F}{}{} terms in Fig.
\ref{bfac_co}e and \ref{bfac_co}f.

\subsection{NLTE and HFS effects on the line formation}{\label{sec:nlte}}

The computed profiles for selected lines of \ion{Co}{i} and \ion{Co}{ii} are
compared with the solar spectrum in Fig. \ref{ltenlte_co}. The synthetic 
profiles were calculated under LTE and NLTE conditions, with and without HFS.

All lines selected for the abundance analysis in the Sun and metal-poor stars
are relatively weak (Table \ref{line_data_coi}). Hence, the \ion{Co}{i} lines
computed under NLTE behave similarly: they are uniformly weakened compared to
the LTE profiles. The main effect is the shift of the optical depth scale due to
$b_i < 1$ at line formation depths. As a result, the NLTE abundance 
corrections\footnote{The difference in abundances required to fit NLTE and LTE
profiles is referred to as the NLTE (abundance) correction $\Delta_{\rm NLTE} =
\log\varepsilon^{\rm NLTE} - \log\varepsilon^{\rm LTE}$.} $\Delta_{\rm NLTE}$ 
are positive; for the majority of \ion{Co}{i} lines computed with the solar
model atmosphere, $\Delta_{\rm NLTE} \sim +0.15$ dex.

NLTE abundance corrections for different stellar parameters are given in Table
\ref{grid_co}. These values are computed only for five \ion{Co}{i} lines and
one \ion{Co}{ii} line, which can be detected in our spectra of stars with
[Fe/H]$ <-0.5$. Although it is not our goal to investigate giant stars, we have
performed test calculations for the \emph{cool giant model} ($\Teff = 4800$ K,
$\log g = 1.8$, [Fe/H] $= -3.3$). This exception is made because many analyses
of Co abundances refer to bright cool giants with low [Fe/H]. NLTE corrections
from Table \ref{grid_co} should only be used to get a crude estimate of
deviations from LTE in metal-poor stars due to the dependence of NLTE correction
on the Co abundance.

\begin{table}
\begin{center}
\caption[]{NLTE abundance corrections for lines of \ion{Co}{i} and \ion{Co}{ii}
calculated with selected models of the grid. Hyphens refer to the lines with
computed NLTE equivalent widths below $1$ \mA. Collisions with neutral hydrogen
are included with the reference scaling factor $S_\mathrm{H} = 0.05$. Note that
$\Delta_\mathrm{NLTE}$ for the \ion{Co}{ii} line at $3501$ \AA\ are also given
for the cases when its equivalent width is below $1$ \mA.}
\vspace{5mm}
\label{grid_co}
\begin{tabular}{l|l|ll|lll}
\hline
$T_\mathrm{eff}$/$\log g$/[Fe/H] & \multicolumn{6}{c}{$\Delta_\mathrm{NLTE}$} \\
\cline{2-7}
\multicolumn{1}{c}{wavelength, $\AA$} & \multicolumn{1}{c}{3501} &
\multicolumn{1}{c}{3845}
& \multicolumn{1}{c}{3957} & \multicolumn{1}{c}{4020} & \multicolumn{1}{c}{4110}
& \multicolumn{1}{c}{4121}\\
\hline
4800/1.8/-3.3 & ~~0.12 & 0.92 & 0.82 &  -   & 0.88 & 0.88  \\
5000/3/ 0 & --0.03 & 0.07 &  0.11 & 0.11 & 0.15 & 0.07  \\
5000/3/-1 & --0.03 & 0.20 &  0.26 & 0.25 & 0.28 & 0.22  \\
5000/3/-2 & ~~0.02 & 0.64 &  0.60 & 0.64 & 0.48 & 0.76  \\
5000/3/-3 & ~~0.13 & 1.03 &  0.85 &  -   & 0.69 & 0.85  \\
5000/4/ 0 & --0.04 & 0.03 &  0.05 & 0.08 & 0.15 & 0.04  \\
5000/4/-1 & --0.04 & 0.13 &  0.17 & 0.18 & 0.28 & 0.15  \\
5000/4/-2 & --0.03 & 0.45 &  0.44 & 0.47 & 0.48 & 0.62  \\
5000/4/-3 & ~~0.08 & 0.67 &   -   &  -   & 0.69 & 0.66  \\
5500/4/ 0 & --0.03 & 0.08 &  0.12 & 0.12 & 0.23 & 0.1   \\
5500/4/-1 & --0.01 & 0.27 &  0.3  & 0.32 & 0.36 & 0.35  \\
5500/4/-2 & ~~0.02 & 0.66 &   -   & 0.63 & 0.63 & 0.67  \\
5500/4/-3 & ~~0.17 & 0.75 &   -   &  -   &  -   & 0.74  \\
5780/4.4/ 0   & --0.02 & 0.11 &  0.14 & 0.12 & 0.23 & 0.1  \\
6000/4/ 0 & --0.02 & 0.1  &  0.15 & 0.18 & 0.22 & 0.13  \\
6000/4/-1 & ~~0.01 & 0.4  &  0.39 & 0.39 & 0.4  & 0.47  \\
6000/4/-2 & ~~0.06 & 0.64 &   -   &  -   & 0.63 & 0.63  \\
6000/4/-3 & ~~0.31 & 0.69 &   -   &  -   &  -   & 0.7   \\
6200/3.4/ 0   & --0.02 & 0.15 &  0.17 & 0.18 & 0.23 & 0.16  \\
6200/3.4/-1.2 & ~~0.01 & 0.5  &  0.47 & 0.46 & 0.47 & 0.52  \\
6200/3.4/-2.4 & ~~0.11 & 0.65 &   -   &  -   &  -   & 0.64  \\
6200/4.6/ 0   & --0.01 & 0.1  &  0.13 & 0.13 & 0.17 & 0.12  \\
6200/4.6/-1.2 & ~~0.01 & 0.43 &  0.42 & 0.41 & 0.42 & 0.46  \\
6200/4.6/-2.4 & ~~0.11 & 0.61 &   -   &  -   &  -   & 0.59  \\
6400/4.2/ 0   & --0.01 & 0.15 &  0.17 & 0.18 & 0.2  & 0.15  \\
\hline
\end{tabular}
\end{center}
\end{table}

\begin{figure}
\resizebox{\columnwidth}{!}{\rotatebox{-90}{\includegraphics{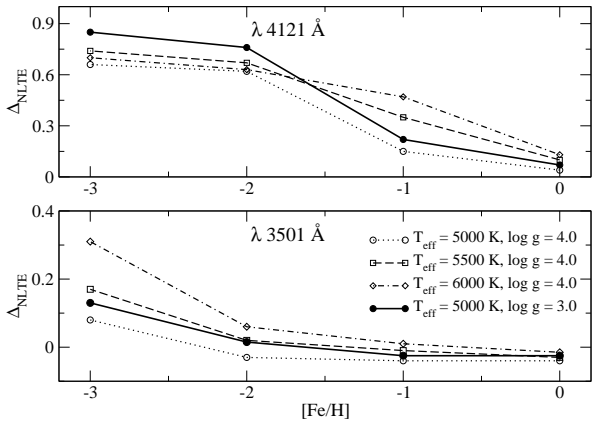}}}
\hfill
\caption{NLTE abundance corrections $\Delta_\mathrm{NLTE}$ calculated for the
lines of \ion{Co}{i} ($4121$ \AA, top) and \ion{Co}{ii} ($3501$ \AA, bottom).
Calculations are performed for sixteen models with $\Teff = 5000, 5500, 6000$ K,
$\log g = 3, 4$, $\mathrm{[Fe/H]} = 0,-1,-2, -3$.}
\label{co_nltecor_gr4}
\end{figure}

The main stellar parameter that determines the sign and magnitude of the NLTE
abundance corrections is metallicity. Low metal abundances lead to decreasing
number density of free electrons and reduced line absorption in the UV, so that
the effect of overionization on the \emph{opacity} of \ion{Co}{i} lines
monotonously increases. On average, $\Delta_{\rm NLTE}$ increases with 
decreasing [Fe/H] from $\sim 0.1$ dex for [Fe/H] $= 0$ to $\sim 0.7$ dex for
[Fe/H] $= -3$ (Fig. \ref{co_nltecor_gr4} top). It is clear that in addition to
the effect of metallicity, high temperatures control overionization at higher
[Fe/H], whereas at low [Fe/H] the effect of low $\log g$ is more important.

The response of NLTE corrections for \ion{Co}{i} to the effective temperature is
not uniform. At moderate gravities ($\log g = 4$), the NLTE corrections are
maximal in the warm model $\Teff = 5500$ with the smallest metallicity (see
Table \ref{grid_co}). Lower NLTE corrections at even higher temperatures are
due to the increased rates of collisions. At low temperatures, NLTE abundance
corrections are also smaller because the stellar flux maximum is shifted to
longer wavelengths, away from ionization thresholds of the important \ion{Co}{i}
levels.

NLTE corrections increase with decreasing surface gravity. However, for
different metallicities and \emph{supersolar} temperatures, reduction of $\log
g$ from $4.6$ to $3.4$ changes the NLTE abundance corrections for all lines only
by $\sim 0.03 - 0.06$ dex. In contrast, $\Delta_\mathrm{NLTE}$ strongly depends 
on stellar gravity in the \emph{cool} model with $\Teff \leq 5000$. The extreme 
values of $\Delta_\mathrm{NLTE}$ are found for the cool giant model with a
subsolar temperature $\Teff = 4800$ K, [Fe/H] $= -3.3$, and a very low gravity
$\log g = 1.8$. This is an expected result, because all collisional interactions
become very weak due to the small metallicity and gravity, and the role of
radiative processes increases in spite of the reduced temperature.

The NLTE corrections for the \ion{Co}{ii} line show a different behaviour, which
is related to the deviation of line source function $S^{\rm l}$ from the Planck
function $B_{\nu}(\Te)$. As an example, we study the formation of the
\ion{Co}{ii} line at $3501$ \AA. It \emph{seems} to be relatively unblended in
the solar spectrum and can be used to test the ionization equilibrium of Co in
the metal-poor stars. In the solar model, this line is sufficiently strong; its
core is formed in the upper layers $\opd \sim -2$, where spontaneous transitions
depopulate the upper level, thus $b_i > b_j$ and $S^{\rm l}~ < ~B_{\nu}(\Te)$.
Therefore, line core intensities are slightly larger in NLTE (Fig.
\ref{ltenlte_co} bottom), and $\Delta_\mathrm{NLTE} < 0$ (Fig.
\ref{co_nltecor_gr4} bottom). At [Fe/H] $ \leq -1$, NLTE abundance corrections
start to increase smoothly. Due to less UV opacity, the radiation field is
amplified producing overpopulation of the higher levels via the photon pumping
mechanism, $b_i < b_j$. The line source function becomes superthermal, $S^{\rm
l}~ > ~B_{\nu}(\Te)$, and the line is weakened compared to the LTE case. The
same mechanism is responsible for the positive NLTE abundance corrections in
the models with [Fe/H] $\leq -2$. At very low metallicities (Fig.
\ref{co_nltecor_gr4} bottom), NLTE corrections are as large as $+0.2 \ldots
+0.3$ dex, depending on the temperature. Consequently, \emph{the analysis of
\ion{Co}{ii} lines in very metal-poor stars can lead to substantial errors in
abundances if NLTE effects are neglected}.

The errors in abundances introduced by neglecting HFS depend on the line
strength. For the solar model, the abundances derived for saturated lines $4121$
\AA\ and $3501$ \AA\ without HFS are overestimated by $0.4$ dex and $0.9$ dex,
respectively. The reason is that HFS effectively de-saturates strong lines
leading to a compound profile, where the strength of each component is
more linearly proportional to the element abundance. Note also that saturated
lines are also sensitive to the $A$ and $B$ factors used in calculations of HFS.
For example, the $3501$ \AA\ HFS line computed with an approximate estimate $A =
49$ mK for the lower level \Co{a}{5}{P}{}{3} \citep[taken
from][]{1998ApJS..117..261P} cannot fit the observed profile at all (Fig.
\ref{ltenlte_co} bottom). In contrast, our new reference value $A = 40$ mK
provides an excellent fit to the observed profile. The weak lines, like $6082$
\AA, are less affected: the profile computed with all HFS components gives $\sim
0.1$ dex lower abundance. At low metallicity, [Fe/H] $= -2$, the abundances are
almost insensitive to inclusion of HFS in spectrum synthesis; the errors for all
investigated lines are not larger than $0.05$ dex.
\subsection{Solar abundance of Co}

The solar spectrum is calculated using the MAFAGS-ODF model atmosphere with
$T_{\rm eff} = 5780$ K, $\log g = 4.44$, [Fe/H] $= 0$, and a constant
microturbulence velocity $\xi_{\rm t} = 0.9$ \kms\ for all lines. The Co
line profiles are broadened by a rotation velocity $V_{\rm rot} = 1.8$ \kms, and
by a macroturbulence velocity $V_{\rm mac}= 2.5 \ldots 4$ \kms. The observed
flux spectrum is taken from the Kitt Peak Solar Flux Atlas
\citep{1984sfat.book.....K}.

The lines of \ion{Co}{i} and \ion{Co}{ii} selected for abundance calculations 
are listed in Table \ref{line_data_coi}. Measured oscillator strengths are 
available only for $17$ out of $20$ lines. Two reference sources of
experimental $\log gf$ values are \citet{1982ApJ...260..395C} and
\citet{1999ApJS..122..557N}. For three yellow lines of multiplet $158$ we had to
adopt the theoretical values of \citet{1996yCat.6010....0K}. However, the
accuracy of calculated data is unknown, hence the lines of multiplet $158$ are
only used in the differential abundance analysis of metal-poor stars. Our final
estimate of the solar Co abundance is based solely on the experimental $\log gf$
values. Three lines in the near-UV, $3845$, $3957$, and $4066$ \AA, are also
neglected in solar calculations due to severe blending.

\begin{figure}
\resizebox{\columnwidth}{!}{\rotatebox{-90}
{\includegraphics{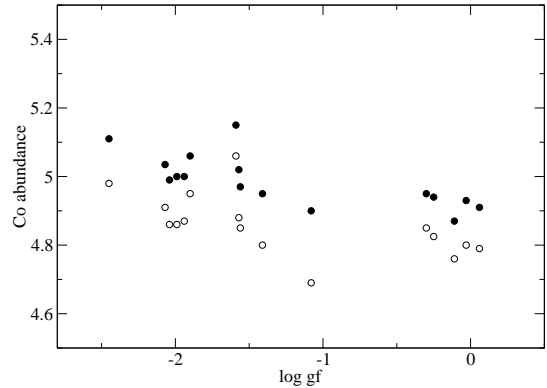}}}
\caption{Abundances of the 17 solar \ion{Co}{i} lines. NLTE and LTE abundances
are marked with filled and open circles, respectively.}
\label{abundall_co}
\end{figure}

Figure \ref{abundall_co} shows NLTE and LTE abundances for $16$ lines of 
\ion{Co}{i} as a function of their oscillator strengths. A weak trend of 
individual abundances with $gf$ values, as well as similar line-to-line scatter 
under NLTE and LTE, points at the errors in $\log gf$'s. For the weak lines, 
the errors are $\sim 15 - 25 \%$ \citep{1982ApJ...260..395C}. Two weak lines at 
$6189$ and $6814$ \AA\ show NLTE and LTE abundances deviating from the mean by 
$\geq 0.1$ dex. The lines at $7417$ and $7712$ \AA\ contain a well-resolved 
blend in their wings, which is, however, not present in our linelist. The line 
at $5369$ \AA\ gives very large abundance, $5.15$ dex. As its $gf$-value seems 
to be reliable, it is likely that the line contains an unidentified blend. We 
ignore these five lines in the solar analysis, and our sample is further
reduced to $11$ lines. No significant correlation of individual NLTE or LTE 
abundances is found with the excitation potential of the lower level or
equivalent width of the lines.

The average NLTE and LTE abundances of Co in the solar atmosphere are $\logeCoN\
= 4.95 \pm 0.04$ dex and $\logeCoL\ = 4.81 \pm 0.05$ dex, respectively. The
standard deviations are quoted as errors. When calculated only with the
$\log gf$ data from \citet{1982ApJ...260..395C}, $\logeCoN\ = 4.96 \pm 0.06$
dex. This value is nearly equal to the NLTE abundance calculated using $5$ lines
with oscillator strengths from \citet{1999ApJS..122..557N}, $\logeCoN\ = 4.98
\pm\ 0.03$.

It is interesting that there is a large difference of $0.14$ dex between
the solar NLTE and LTE abundances determined from the \ion{Co}{i} lines with
$\SH = 0.05$. Up to now, such a strong NLTE abundance effect has been
demonstrated only for \ion{Sc}{i}, with $\Delta_{\rm NLTE} \sim 0.18$ dex
derived using $\SH = 0.1$ \citep{2008A&A...481..489Z}. The average solar NLTE
corrections for the other Fe-peak elements, Fe \citep{2001A&A...380..645G} and
Mn \citep{2007A&A...473..291B}, are smaller. This is, however, not surprising
because the magnitude of NLTE effects strongly depends on the \emph{model atom
used in calculations}, in particular, on the collision efficiency with hydrogen
atoms $\SH$ (see Sect. \ref{sec:atmmod}) and the detailed atomic properties,
like separation of energy levels, number and strength of transitions,
and ionization potential. Weaker NLTE effects for \ion{Mn}{i} ($\Delta_{\rm
NLTE} \sim 0.05$ dex) compared to \ion{Co}{i} are due to the differences in the
atomic models: a) the number of low-excitation levels, which stipulate
overionization; b) collisional coupling to the continuum \citep[the latter issue
was also emphasized in][]{2001A&A...380..645G}, which amplifies recombination;
c) the general number of levels. The model of the \ion{Fe}{i} atom used by
\citet{2001A&A...366..981G} is different from our \ion{Co}{i} model (at least)
in the following aspects. The authors adopted $\SH = 5$ as their final value
\emph{and} forced thermalization of the \ion{Fe}{i} levels with excitation
energy above $7.3$ eV. Assuming \emph{no} explicit thermalization, as we do in
the current work, \citet{2001A&A...366..981G} obtain \emph{very strong NLTE
effects} on \ion{Fe}{i} lines even for the model with $\SH = 5$, $\Delta_{\rm
NLTE} \sim 0.12$ dex. Using this large scaling factor for \ion{H}{i} collisions,
we derive the average NLTE abundance correction for \ion{Co}{i} lines
$\Delta_{\rm NLTE} \sim 0.06$ dex. Thus, we see no reason to force
thermalization of the upper \ion{Co}{i} levels. In contrast, we consider
collisions with \ion{H}{i} as the main source of $\logeCoN\ $ uncertainty. We
performed a series of NLTE abundance calculations with $0 \leq \SH \leq 5$. The 
NLTE result for $\SH = 0$, $\logeCoN\ = 4.96 \pm 0.04$, is almost identical to
the case of $\SH = 0.05$. Also, with increasing scaling factor from $0.5$ to
$5$, the abundance steadily decreases from $4.91$ to $4.87$ dex with the
standard deviation remaining constant, $\sigma \approx 0.04$ dex. Thus, no
conclusion can be drawn concerning the optimum value of the scaling factor to
hydrogen collisions.

The majority of \ion{Co}{i} lines are weak, and they are not sensitive to the 
van der Waals damping, neither are they affected by the microturbulence. We 
found only one line at $4121$ \AA, which gives different abundances with the
variation of $\Vmic$ or $\log C_6$. For $\Delta \Vmic = \pm 0.2$ \kms, $\Delta
\loge = \mp 0.1$ dex; $\Delta \log C_6 = \pm 0.5$ corresponds to $\Delta \loge =
\mp 0.12$ dex. With $\log C_6 = -31.63$ calculated using
\citet{1995MNRAS.276..859A} theory and $\Vmic = 0.9$ \kms, this line gives a
NLTE abundance of $4.95$ dex, which is consistent with abundances derived from
the other \ion{Co}{i} lines. Moreover, the standard deviation of the
\emph{average} Co NLTE abundance, $\sigma = 0.04$ dex, is already smaller than
the errors given for the oscillator strengths. Hence, we have not tried to
adjust damping constants and microturbulence any further.

We checked the \emph{ionization equilibrium} of Co for the Sun, using the
relatively unblended line of ionized cobalt \ion{Co}{ii} at $3501$ \AA\ (Fig.
\ref{ltenlte_co} bottom). The line is saturated and consists of $8$ HFS
components, which were calculated with our new hyperfine splitting constants. To
produce a better fit to the observed profile, the $A$ factor for the upper level
was decreased by $1$ mK; this is within the errors of the measured HFS value.
Using the theoretical oscillator strength from \citet{1998A&AS..130..541R},
$\log gf = -1.22$, we derive $\logeCoN\ = 4.86$ and $\logeCoL\ = 4.88$ dex. With
the experimental $gf$-value from \citet{1985PhRvA..31..744S}, $\log gf = -1.18$,
the NLTE and LTE abundances are $\logeCoN\ = 4.82$ and $\logeCoN\ = 4.84$ dex,
respectively. We do not assign any error to these values for the reasons
described in the following paragraph.

The $\sim 0.1$ dex discrepancy between the lines of \ion{Co}{i} and \ion{Co}{ii}
in NLTE is most likely due to the erroneous abundance from the \ion{Co}{ii}
line. First, the uncertainty due to the continuum placement and blending in the
near-UV window of the solar spectrum is large. At present, there is no
information about other strong lines overlapping with the $3501$ \AA\ line of
\ion{Co}{ii}. However, if atomic physics experiments or theoretical calculations
reveal the presence of a strong blend, the Co abundance determined from the UV
\ion{Co}{ii} line will be reduced. The accuracy of the calculated oscillator
strength provided by \citet{1998A&AS..130..541R} is not known, whereas the
accuracy of the \citet{1985PhRvA..31..744S} $gf$-value for the $3501$ \AA\ line
is not better than $50\%$, which translates to abundance uncertainty of $\sim
0.2$ dex. If the problem is concealed in hydrogen collisions, a very large
scaling factor $\SH \geq 5$ is needed to bring two ionization stages in
agreement. This issue will be further investigated in the next section devoted
to the metal-poor stars.

The abundance of Co in \ion{C}{i} meteorites is $4.89 \pm 0.01$ dex
\citep{2009arXiv0901.1149L}. It conforms to our NLTE abundance $\logeCoN\ = 4.95
\pm 0.04$ dex derived from \ion{Co}{i} lines within the combined errors of both
values. Also, the LTE abundance from the \ion{Co}{ii} line ($4.88$ dex) agrees
to meteoritic, if the theoretical $gf$-value from \citet{1998A&AS..130..541R} is
used. Previously, the photospheric LTE abundance from the \ion{Co}{i} lines was
determined by \citet{1982ApJ...260..395C}, $\logeCo\ = 4.92 \pm 0.08$ dex. This
value is $\sim 0.1$ dex higher than our LTE abundance, which is due to several
reasons. First, the authors used the semi-empirical model atmosphere of
\citet{1967ZA.....65..365H}, which is the essentially the same as the model of 
\citet{1974SoPh...39...19H}. The latter model is known to give higher abundances
compared to the theoretical model atmospheres. Second, equivalent widths were
used to measure the abundances. This method always gives only an \emph{upper}
limit on the abundance, because the contribution to a line strength from
blending lines is neglected. In all other respects (oscillator strengths,
microturbulence, line set), our study and the analysis of
\citet{1982ApJ...260..395C} are similar.

\subsection{Abundances in metal-poor stars}

Our sample of $17$ objects from \citet{2008PhST..133a4013B} was expanded by one
subdwarf G 64-12 with [Fe/H] = $-3.12$. The UVES spectrum for G 64-12 was taken
from ESO/ST-ECF Science Archive Facility (PID 67.D-0554(A)). Other stars were
observed with the UVES echelle spectrograph (ESO VLT UT2, Chile) in 2001, and/or
with the FOCES echelle spectrograph (CAHA observatory, Calar Alto) during 1999
and 2000. The data for HD 61421 and HD 84937 were taken from the UVESPOP survey
\citep{2003Msngr.114...10B}. The UVES data cover a spectral range from $3\,300$
to $6\,700$ \AA, whereas FOCES spectra are cut at $4\,000$ \AA. A detailed
description of the observational material can be found in
\citet{2008PhST..133a4013B} and \citet{2008A&A...492..823B}.
\begin{figure}
\resizebox{\columnwidth}{!}{\rotatebox{-90}
{\includegraphics{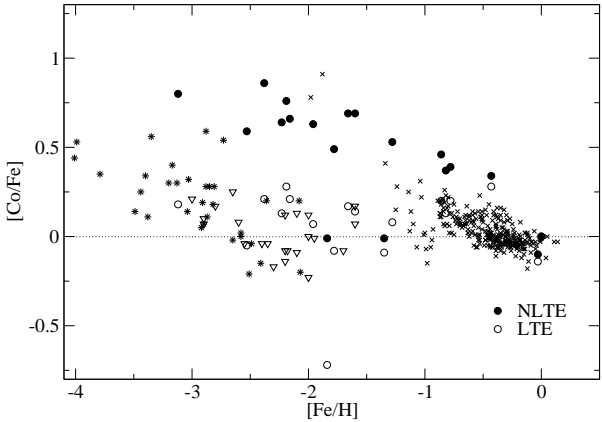}}}
\caption{[Co/Fe] vs. [Fe/H] in metal-poor stars. The Co abundances are derived
from \ion{Co}{i} lines under NLTE (filled circles) and LTE (open circles).
Iron abundances are determined from \ion{Fe}{ii} lines in LTE. The sources of
other LTE [Co/Fe] data are: \citealt{2002ApJS..139..219J} (open triangles),
\citealt{2003MNRAS.340..304R,2006MNRAS.367.1329R} (crosses),
\citealt{1995AJ....109.2757M} (stars).}
\label{ab_co_1}
\end{figure}

The abundance ratios for Co with respect to Fe in stars of our sample are given
in Table \ref{stel_param_gen}. The Co abundances of thick disk stars with $-1 <$
[Fe/H] $< -0$ are based on $6$ to $10$ lines, which have $W_{\lambda} > 20$ \mA\
in the solar spectrum. In the spectra of the halo stars, only a few lines of
\ion{Co}{i} ($\lambda\lambda\ 3845$, $4020$, $4110$, $4121$) are detected. Due
to a systematic abundance offset of $\sim 0.2$ dex between the line at $4110$
\AA\ and other \ion{Co}{i} lines, this near-UV line was excluded from stellar
analysis. The discrepancy might be related to the poor modelling of H$_\delta$
inner wings, where the $4110$ \AA\ line is located, or to a blend. In fact,
no Co abundance studies in metal-poor stars include the line at $4110$ \AA. For
G 64-12, the abundance is based on the single \ion{Co}{i} line at $3845$ \AA,
and the error bar $\pm 0.1$ dex reflects the uncertainty of the line fitting due
to the low S/N (S/N $\sim 120$) in the near-UV spectral window. We could not
determine the solar $\loggfesun$ value for the $3845$ \AA\ line, hence all
stellar abundances based on the line were calculated using the experimental
$\log gf = 0.01$ from \citet{1982ApJ...260..395C}.

The [Co/Fe] - [Fe/H] relation, along with the results of other authors, is
demonstrated in Fig. \ref{ab_co_1}. Note that the solar NLTE and LTE [Co/Fe]
values are set to $0$ by definition. The NLTE abundances of Co in the metal-poor
stars are higher than the LTE abundances by $0.4 - 0.7$ dex. Thus, instead of
the flat [Co/Fe] trend in metal-poor stars, we find that the [Co/Fe] ratio
increases with decreasing metallicity down to [Fe/H] $\sim -2$ and is roughly
constant thereafter. Two stars with very low NLTE and LTE [Co/Fe] ratios are HD
25329 and HD 103095. These stars were also analysed by
\citet{1981ApJ...244..989P}, who derived [Co/Fe] $= +0.48$ and [Co/Fe] $= +0.15$
for HD 25329 and HD 103095, respectively. The discrepancy with our values stems
from different model parameters and abundance analysis techniques. In
particular, Peterson used $\log gf$ values calibrated on solar lines to
calculate the abundances from line equivalent widths, and Uns\"old approximation
for $\log C_6$ values was assumed.

Our LTE abundances are generally consistent with LTE results of the other
authors considered here, although in the metal-poor domain Co abundances show a
large scatter within a study and between studies (Fig. \ref{ab_co_1}). The
general trend is hard to define, although the majority of studies indicate that
Co is increasing with respect to iron towards very low metallicities. For the
thick disk stars our [Co/Fe] trend with [Fe/H] is flat and slightly supersolar,
in agreement with \citet{2006MNRAS.367.1329R}. The thin disk star, Procyon,
supports mild underabundance of Co in the thin disk as follows from the results
of \citet{2003MNRAS.340..304R}. The average Co abundances for stars [Fe/H] $<
-1$ are consistent with \citet{2002ApJS..139..219J} as to slope, absolute
values, and dispersion. Note that Johnson analyzed only giants.

In the attempt to check the ionization equilibrium of Co, we calculated
differential abundances from the $3501$ \AA\ line of \ion{Co}{ii}. This line can
still be discerned from the continuum only in the spectra of $5$ stars. Thus,
these abundances must be treated, at best, as upper limits because of poor S/N
below $4000$ \AA. The NLTE abundance corrections for \ion{Co}{ii} are not so
small and positive. For HD 84937 and HD 140283, $\Delta$[Co/Fe]$_{\rm II}$(NLTE
- LTE)\footnote{Co abundance ratios determined from the \ion{Co}{i} and
\ion{Co}{ii} lines are denoted as [Co/Fe]$_{\rm I}$ and [Co/Fe]$_{\rm II}$,
respectively}$ = +0.14$ and $+0.11$ dex, respectively. Whereas, for HD 102200,
HD 34328, and HD 122563, the NLTE corrections are $+0.04 \ldots +0.07$ dex.
A significant discrepancy still remains between abundances determined from the
lines of \ion{Co}{i} and \ion{Co}{ii}.

Consistent ionization equilibria under NLTE require either significantly lower
abundances from the neutral lines and/or much higher abundances from the ionic
line. The former can be achieved with increasing the scaling factor to the
\ion{H}{i} collision cross-sections, $\SH$. Test calculations were performed for
two stars. The NLTE results for HD 34328 computed with $\SH = 1$ are not very
different from the case of $\SH = 0.05$ (our reference value): the abundance
from the \ion{Co}{i} lines decreased by $0.15$ dex. It is only with $\SH = 10$
that the discrepancy is solved: [Co/Fe]$_{\rm I}^{\rm NLTE} = 0.26$, which is
in agreement with [Co/Fe]$_{\rm II}^{\rm NLTE} = 0.21$. The abundances derived
from \ion{Co}{ii} lines are not sensitive to a variation of $\SH$. However,
for HD 84937, assumption of $\SH = 10$ leads to [Co/Fe]$_{\rm I} = 0.27$, which
is still $\approx 0.2$ dex higher than the NLTE abundance determined from
\ion{Co}{ii} lines. Whether such a large scaling factor is justified is
questionable.

Another option is that our model atmospheres are not satisfactory, and granular
inhomogeneities will have a large effect on abundances in addition to NLTE.
Let us consider Fe, for which calculations of 1.5D NLTE line formation with 3D
convective model atmospheres have been performed for the Sun and the metal-poor
subgiant HD 140283. Using the \ion{Fe}{i} lines \citet{2005ApJ...618..939S}
found equal 3D and 1D NLTE abundances for both stars that were interpreted as
due to \emph{stronger} NLTE effects in 3D than in 1D models, which exactly
cancel large 3D corrections. However, large positive NLTE abundance corrections
of $\sim +0.4$ dex were demonstrated at low metallicity for weak \ion{Fe}{ii}
lines in $3$D. Recalling that \ion{Co}{i} and \ion{Fe}{i} have a very similar
complex atomic structure with ionization energies differing by $\sim 0.04$ eV,
one may expect that 3D will have the same (minor) effect on NLTE-based
abundances of Co determined from \ion{Co}{i} lines. Whereas, \emph{large
positive 3D corrections may be necessary for \ion{Co}{ii} lines}. In this case,
our metallicities based on LTE abundances from \ion{Fe}{ii} lines may not
necessarily be wrong, because the magnitude of the NLTE effect depends strongly
on the choice of the scaling factor $\SH$ to the hydrogen collision
cross-sections. This parameter must be separately calibrated on the solar and
stellar spectra, and for Fe it may take the values different from that used for
Co.

We have also investigated the effect of a temperature enhancement on the 
abundance analysis by repeating the calculations for HD 102200 with $\Delta
\Teff = \pm 100$ K with respect to our reference value $\Teff = 6120$ K. For
both ions the effect turned out to be negligible. With $\Teff = 6220$ K,
abundances for \ion{Co}{i} lines increased respectively by $0.02$ and $0.06$
dex, and the abundances from the \ion{Co}{ii} line decreased by $\sim 0.02$ dex.
Abundance corrections of the same magnitude but with the reverse sign were
derived for lower $\Teff$. Thus, we cannot attribute the ionization equilibrium
problem to systematic errors in $\Teff$.

\section{Comparison with chemical evolution models}

There are several models in the literature that describe the evolution of Co
in the Galaxy. The difference between theoretical trends is strikingly large:
both the shape of the [Co/Fe] trend and the magnitude of Co depletion or
overproduction relative to Fe in metal-poor stars vary by an order of magnitude.
In fact, \emph{neither of the GCE models are able to describe our NLTE [Co/Fe]
trend in metal-poor stars}. The \citet{2000A&A...359..191G} model with
metallicity-independent SNe yields gives only a qualitatively similar behaviour,
characterized by increase of [Co/Fe] ratios towards low metallicities. However,
this model gives a very good fit to the LTE Co abundances from the current work
and other sources. Closer to our NLTE results is the model of
\citet{2004A&A...421..613F} with unadjusted metallicity-independent SN yields,
but the offset between theoretical calculations and the spectroscopic abundances
is still too large, $\sim 0.25$ dex. The chemical evolution models with
metallicity-dependent SN yields
\citep{1995ApJS...98..617T,1998ApJ...496..155S,2000A&A...359..191G}
predict radically different [Co/Fe] trends, which are ruled out both by NLTE
and LTE spectroscopic abundances of Co. The same is true for the results of
\citet{1971ApJ...166..153A} Co evolution calculations, whether with or
without hydrostatic carbon burning. This discrepancy was also noted by
\citet{1971ApJ...165...87A} and \citet{1971ApJ...166..153A}, who proposed that
the [Co/Fe] ratio should be strongly correlated with the SN Ia progenitor metal
content and the amount of mixing in the ISM. However, the effect that this 
would have on the abundances in metal-poor stars is unclear.

The conclusion from the NLTE trend of Co in metal-poor stars is that massive
stars overproduce Co relative to Fe. The observed decline of [Co/Fe] at [Fe/H]
$\sim -1$ can be explained by an underproduction of Co in SN Ia, but the
possibility that SN II and SN Ia Co yields are metallicity-dependent is also
not excluded. However, we see that all chemical evolution models, which utilize
metallicity-dependent yields, are inadequate to describe the [Co/Fe] trend
suggesting that \emph{the problem is in the stellar yields}. This is not
unexpected, given the sensitivity of supernova yields to the details of
explosion. \citet{1995ApJS..101..181W} emphasized that their SN II iron yields
could be in error by a factor of $2$, ascribing even larger uncertainties to the
elements produced in $\alpha$-rich freeze-out due to the great sensitivity of
those to the mass cut placement. For deeper mass cuts, more Co is produced.
\citet{1999ApJ...517..193N} found that an increasing neutron excess increases
the Co production. \citet{2003Natur.422..871U, 2005ApJ...619..427U} showed that
if a large-scale uniform mixing occurs in the inner layers of very massive
stars, $M > 25 M_{\odot}$, then Co is produced more efficiently and the effect
is even enhanced for larger explosion energies. In connection with this,
\citet{1997ApJ...486.1026N} demonstrated that the axisymmetric explosion in SN
II models also boosts an $\alpha$-rich freeze-out, and hence an increased 
production of Co. Interestingly, the results of massive star nucleosynthesis 
calculations performed by \citet{2008arXiv0803.3161H} suggest that there is no
need for hypernova to explain element abundance patterns in extremely metal-poor
stars. The variety of mechanisms, which could affect the Fe-peak production, is
overwhelming \citep{2007PrPNP..59...74T}. Thus it is not hard to believe that
the disagreement between observations and predictions of chemical evolution
models is, at least in part, due to the deficiencies of the latter.

\section{Conclusions}

Deviations from LTE for \ion{Co}{i} are controlled by overionization in the
whole range of stellar parameters considered in this work. Thus, ionization
equilibrium of Co is shifted towards lower number densities of \ion{Co}{i}
atoms compared to LTE. Excitation equilibrium of \ion{Co}{ii} is also disturbed
due to optical line pumping. 

NLTE effects on the line formation occur for lines of both ionization stages.
For \ion{Co}{i}, reduced line opacity stipulated by the overionization leads to
general weakening of lines. Lines of \ion{Co}{ii} are affected by the deviation
of line source functions from the Planck function. In the solar model
atmosphere, these deviations are small, and it is safe to adopt LTE for
\ion{Co}{ii}.  In the models with very low metallicities and supersolar
temperatures, NLTE corrections for \ion{Co}{ii} lines are as large as $+0.3$
dex. NLTE abundance corrections for the lines of \ion{Co}{i} vary from $+0.1$ to
$+0.6$ dex depending on the metallicity and the effective temperature. Our
results for the giant models with low $\Teff$, low $\log g$, and low [Fe/H]
suggest that in atmospheres of giants deviations from LTE are \emph{larger} than
in dwarfs, although the main stellar parameter that controls the magnitude of
NLTE effects in Co is metallicity.

The solar abundance of Co derived from the NLTE analysis of \ion{Co}{i}
lines is $4.95 \pm 0.04$ dex, which is $0.13$ dex higher than the LTE abundance.
The abundance calculated from the single relatively unblended \ion{Co}{ii} line
in the solar spectrum is $4.86$ dex. The discrepancy of $0.09$ dex between lines
of two ionization stages points either at a failure of the NLTE approach to
describe the ionization equilibrium of Co, or to an erroneous theoretical
oscillator strength for the \ion{Co}{ii} transition. The NLTE abundances
calculated with different scaling factors to cross-sections for \ion{H}{i}
collisions, $0.05 \leq \SH \leq 5$, are in agreement with the Co abundance
measured in \ion{C}{i} meteorites, $4.89 \pm 0.01$ dex.

For the \emph{solar-metallicity} stars, the incorrect treatment of hyperfine
structure leads to large errors in abundances. For the lines with saturated
cores, the neglect of HFS leads to an abundance overestimate by $> 0.5$ dex.
Even the weak lines in the solar spectrum are not free from HFS effects, showing
differences of $0.05 \ldots 0.15$ dex. The influence of HFS on the line profiles
and abundances is negligible in the metal-poor stars; not exceeding $0.05$ dex.

The differential abundance analysis of $17$ stars with $-2.5 \leq$ [Fe/H] $\leq
0$ demonstrated that the evolution of Co abundances in the Galaxy is
radically different from all previously reported trends. The NLTE abundance
ratios [Co/Fe] are supersolar for the thick disk stars, and at low metallicities
[Co/Fe] are as large as $0.8$ dex. This result suggests that Co is overproduced
relative to Fe in short-lived massive stars. The GCE models, which utilize
metallicity-dependent yields for SNe, predict a qualitatively different [Co/Fe]
trend for the metal-poor stars, characterized by a plateau or [Co/Fe] depletion.
We relate the discrepancy to the deficiency in SN yields.

\section*{Acknowledgments}

We appreciate the assistance of Dr. Paul Barklem in calculations of $\log C_6$
values for the investigated lines. This work is based on observations made with
the European Southern Observatory telescopes obtained from the ESO/ST-ECF 
Science Archive Facility. JCP thanks Ben Earner and Heather Cheeseman who 
contributed to this study through their Master's project. JCP thanks STFC 
(PPARC of the UK) and The Leverhulme Trust for financial support for this work.
We thank the referee for useful comments and corrections.

\bibliographystyle{mn2e}
\bibliography{references}

\appendix
%------------------------------
\section{for online publication}
\begin{figure*}
\centering
\resizebox{\textwidth}{!}{\includegraphics{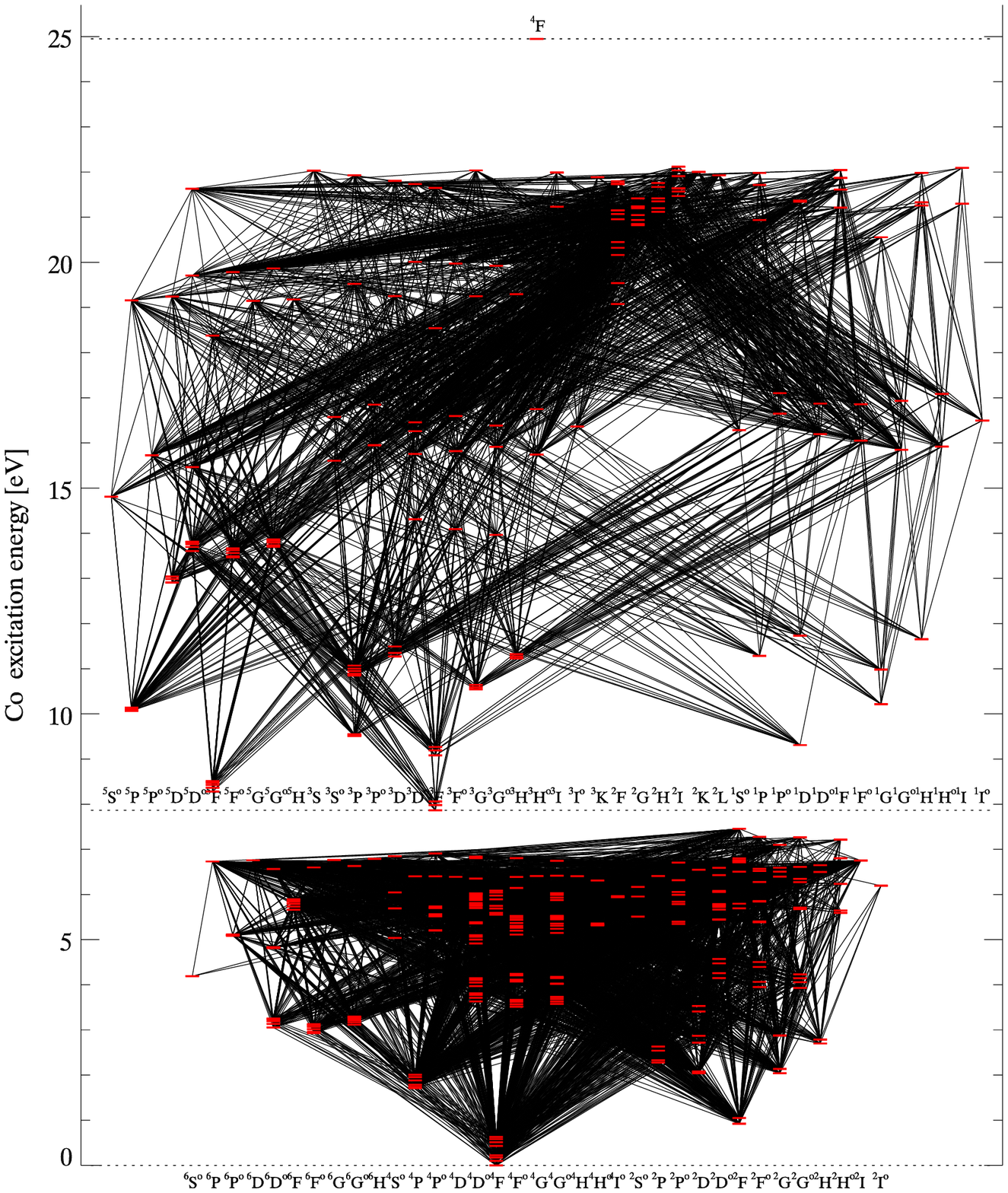}}
\caption[]{Grotrian diagram of the \ion{Co}{i}/\ion{Co}{ii} model atom.
Solid lines represent allowed and forbidden transitions included in the model
atom. Ionization energies of \ion{Co}{i} and \ion{Co}{ii} are $7.86$ eV and
$17.1$ eV, respectively.}
\label{atom}
\end{figure*}

\newpage
\begin{table*}
\begin{minipage}{\linewidth}
\renewcommand{\footnoterule}{}
\renewcommand{\tabcolsep}{2.7mm}
\caption[HFS constants $A$ and $B$ for \ion{Co}{i} and \ion{Co}{ii} levels]{HFS
constants $A$ and $B$ (in units of $10^{-3}$ cm$^{-1}$) representing magnetic
dipole and electric quadrupole interactions for \ion{Co}{i} and \ion{Co}{ii}
levels. Level energies $E$ are given in eV. The interaction constants for all
\ion{Co}{i} levels are taken from Pickering (1996). For the \ion{Co}{ii} levels,
the data measured in this work are used.}
\label{HFS_co}
\begin{center}
\begin{tabular}{rlrrrc|rlrrrc}
\hline No. &  level &  $g$  &  $E$ & $A$ & $B$  & No. &  level &  $g$  &  $E$ & 
 $A$   &   $B$ \\ \hline
  1 & \Co{a}{4}{F}{}{9/2}  & 10  &  0.00 &   15.~~~&   4.6  & 51 &
\Co{z}{4}{F}{o}{3/2} &  4  &  3.67 &   14.2  &  0.0~~ \\ 
  2 & \Co{a}{4}{F}{}{7/2}  &  8  &  0.10 &   16.4  &   3.2  & 52 &
\Co{z}{4}{G}{o}{11/2} & 12  &  3.58 &   25.8  &  7.~~ \\ 
  3 & \Co{a}{4}{F}{}{5/2}  &  6  &  0.17 &   20.5  &   2.3  & 53 &
\Co{z}{4}{G}{o}{9/2} & 10  &  3.63 &   17.3  &  6.~~ \\ 
  4 & \Co{a}{4}{F}{}{3/2}  &  4  &  0.22 &   34.8  &   2.3  & 54 &
\Co{z}{4}{G}{o}{7/2} &  8  &  3.69 &   15.~~~&  5.~~ \\ 
  5 & \Co{b}{4}{F}{}{9/2}  & 10  &  0.43 &   27.7  & --4.~~~& 55 &
\Co{z}{4}{G}{o}{5/2} &  6  &  3.73 &   13.3  &  5.~~ \\ 
  6 & \Co{b}{4}{F}{}{7/2}  &  8  &  0.51 &   22.3  & --2.6  & 56 &
\Co{z}{4}{D}{o}{7/2} &  8  &  3.63 &   25.1  &  0.~~ \\ 
  7 & \Co{b}{4}{F}{}{5/2}  &  6  &  0.58 &   18.8  & --1.8  & 57 &
\Co{z}{4}{D}{o}{5/2} &  6  &  3.71 &   23.2  &  1.~~ \\ 
  8 & \Co{b}{4}{F}{}{3/2}  &  4  &  0.63 &   10.2  & --2.~~~& 58 &
\Co{z}{4}{D}{o}{3/2} &  4  &  3.78 &   23.5  &  1.~~ \\ 
  9 & \Co{a}{2}{F}{}{7/2}  &  8  &  0.92 &   13.~~~& --5.~~~& 59 &
\Co{z}{4}{D}{o}{1/2} &  2  &  3.81 &   27.5  &  0.~~ \\ 
 10 & \Co{a}{2}{F}{}{5/2}  &  6  &  1.05 &   37.1  & --3.~~~& 60 &
\Co{z}{2}{G}{o}{9/2} & 10  &  3.93 &   16.5  &  0.~~ \\ 
 11 & \Co{a}{4}{P}{}{5/2}  &  6  &  1.71 &   5.9   & --8.~~~& 61 &
\Co{z}{2}{G}{o}{7/2} &  8  &  4.06 &   30.7  &  5.~~ \\ 
 12 & \Co{a}{4}{P}{}{3/2}  &  4  &  1.74 &   10.6  &   4.~~~& 62 &
\Co{z}{2}{F}{o}{7/2} &  8  &  3.95 &   15.~~~&  6.~~ \\ 
 13 & \Co{a}{4}{P}{}{1/2}  &  2  &  1.79 & --23.6  &   0.0  & 63 &
\Co{z}{2}{F}{o}{5/2} &  6  &  4.06 &   34.9  &  0.~~ \\ 
 14 & \Co{b}{4}{P}{}{5/2}  &  6  &  1.88 &   37.4  &   5.~~~& 64 &
\Co{y}{4}{D}{o}{7/2} &  8  &  3.97 &   16.~~~&  5.~~ \\ 
 15 & \Co{b}{4}{P}{}{3/2}  &  4  &  1.96 &   15.4  & --2.~~~& 65 &
\Co{y}{4}{D}{o}{5/2} &  6  &  4.05 &   15.5  &  0.~~ \\ 
 16 & \Co{b}{4}{P}{}{1/2}  &  2  &  2.01 &   57.6  &   0.0  & 66 &
\Co{y}{4}{D}{o}{3/2} &  4  &  4.11 &   19.7  &  0.~~ \\ 
 17 & \Co{a}{2}{G}{}{9/2}  & 10  &  2.04 &   20.5  &   2.~~~& 67 &
\Co{y}{4}{D}{o}{1/2} &  2  &  4.15 &   58.9  &  0.~~ \\ 
 18 & \Co{a}{2}{G}{}{7/2}  &  8  &  2.14 &   28.~~~& --3.2  & 68 &
\Co{y}{4}{F}{o}{9/2} & 10  &  4.07 &    9.9  &  8.~~ \\ 
 19 & \Co{a}{2}{D}{}{3/2}  &  4  &  2.04 &   13.~~~&   0.0  & 69 &
\Co{y}{4}{F}{o}{7/2} &  8  &  4.11 &   17.~~~&  0.~~ \\ 
 20 & \Co{a}{2}{D}{}{5/2}  &  6  &  2.08 &   46.3  &   4.~~~& 70 &
\Co{y}{4}{F}{o}{5/2} &  6  &  4.21 &   20.1  &  0.~~ \\ 
 21 & \Co{a}{2}{P}{}{3/2}  &  4  &  2.28 &   11.2  &   4.~~~& 71 &
\Co{y}{4}{F}{o}{3/2} &  4  &  4.24 &   39.7  &--2.~~~\\ 
 22 & \Co{a}{2}{P}{}{1/2}  &  2  &  2.33 &   20.1  &   0.0  & 72 &
\Co{y}{2}{G}{o}{9/2} &  10  &  4.15 &   14.7  &  0.~~ \\ 
 23 & \Co{b}{2}{P}{}{3/2}  &  4  &  2.54 &   5.5   &   2.~~~& 73 &
\Co{z}{2}{D}{o}{5/2} &  6  &  4.15 &   15.4  &  0.~~ \\ 
 24 & \Co{b}{2}{P}{}{1/2}  &  2  &  2.63 &   17.2  &   0.0  & 74 &
\Co{z}{2}{D}{o}{3/2} &  4  &  4.26 &   46.1  &  2.~~ \\ 
 25 & \Co{a}{2}{H}{}{11/2}  & 12  &  2.70 &   22.6  &   0.0  & 75 &
\Co{y}{2}{D}{o}{5/2} &  6  &  4.48 &   16.4  &  0.~~ \\ 
 26 & \Co{a}{2}{H}{}{9/2}  & 10  &  2.79 &   26.4  &   0.0  & 76 &
\Co{y}{2}{D}{o}{3/2} &  4  &  4.57 &   42.~~~&  0.~~ \\ 
 27 & \Co{b}{2}{D}{}{5/2}  &  6  &  2.72 &   18.5  &  15.~~~& 77 &
\Co{y}{2}{F}{o}{7/2} &  8  &  4.40 &   13.9  &  0.~~ \\ 
 28 & \Co{b}{2}{D}{}{3/2}  &  4  &  2.87 &   30.8  &   4.~~~& 78 &
\Co{y}{2}{F}{o}{5/2} &  6  &  4.50 &   30.4  &  0.~~ \\ 
 29 & \Co{b}{2}{G}{}{9/2}  &  10  &  2.87 &   36.9  &   0.0  & 79 &
\Co{x}{4}{D}{o}{7/2} &  8  &  4.92 &   10.1  &  0.~~ \\ 
 30 & \Co{b}{2}{G}{}{7/2}  &  8  &  2.88 &   8.8   &   0.0  & 80 &
\Co{x}{4}{D}{o}{5/2} &  6  &  5.00 &   11.9  &  1.~~ \\ 
 31 & \Co{z}{6}{F}{o}{11/2} & 12  &  2.93 &   28.5  & --2.~~~& 81 &
\Co{x}{4}{D}{o}{3/2} &  4  &  5.06 &   17.6  &  0.~~ \\  
 32 & \Co{z}{6}{F}{o}{9/2} &  10  &  2.96 &   28.5  &   4.~~~& 82 &
\Co{x}{4}{D}{o}{1/2} &  2  &  5.1 &    55.8  &  0.~~ \\ 
 33 & \Co{z}{6}{F}{o}{7/2} &  8  &  3.02 &   26.4  &   5.~~~& 83 &
\Co{e}{4}{F}{ }{9/2} &  10  &  5.55 &   13.4  &  0.~~ \\ 
 34 & \Co{z}{6}{F}{o}{5/2} &  6  &  3.07 &   24.~~~&   2.~~~& 84 &
\Co{e}{4}{F}{ }{7/2} &  8  &  5.59 &   11.1  &  1.~~ \\ 
 35 & \Co{z}{6}{F}{o}{3/2} &  4  &  3.11 &   20.1  &   0.0  & 85 &
\Co{e}{4}{F}{ }{5/2} &  6  &  5.69 &   18.4  &  0.~~ \\ 
 36 & \Co{z}{6}{F}{o}{1/2} &  2  &  3.13 & --2.1   &   0.0  & 86 &
\Co{e}{4}{F}{ }{3/2} &  4  &  5.75 &   35.5  &  0.~~ \\ 
 37 & \Co{z}{6}{D}{o}{9/2} &  10  &  3.05 &   28.1  &   0.0  & 87 &
\Co{e}{6}{F}{ }{11/2} & 12  &  5.66 &   33.5  &  5.~~ \\ 
 38 & \Co{z}{6}{D}{o}{7/2} &  8  &  3.13 &   27.2  & --3.~~~& 88 &
\Co{e}{6}{F}{ }{9/2} &  10 &  5.73 &   31.5  &  2.~~ \\ 
 39 & \Co{z}{6}{D}{o}{5/2} &  6  &  3.19 &   26.6  & --6.~~~& 89 &
\Co{e}{6}{F}{ }{7/2} &  8  &  5.79 &   28.7  &--1.~~~\\ 
 40 & \Co{z}{6}{D}{o}{3/2} &  4  &  3.23 &   27.~~~&   0.0  & 90 &
\Co{e}{6}{F}{ }{5/2} &  6  &  5.84 &   25.7  &--1.~~~\\ 
 41 & \Co{z}{6}{D}{o}{1/2} &  2  &  3.26 &   33.4  &   0.0  & 91 &
\Co{e}{6}{F}{ }{3/2} &  4  &  5.87 &   19.4  &--1.~~~\\ 
 42 & \Co{z}{6}{G}{o}{13/2} & 14  &  3.12 &   25.3  &   8.~~~& 92 &
\Co{e}{6}{F}{ }{1/2} &  2  &  5.89 & --17.1  &  0.~~ \\ 
 43 & \Co{z}{6}{G}{o}{11/2} & 12  &  3.17 &   23.~~~&   7.~~~& 93 &
\Co{f}{4}{F}{ }{9/2} &  10  &  5.89 &   36.~~~&  5.~~ \\ 
 44 & \Co{z}{6}{G}{o}{9/2} &  10  &  3.22 &   20.6  &   5.~~~& 94 &
\Co{f}{4}{F}{ }{7/2} &  8  &  5.98 &   28.3  &  3.~~ \\ 
 45 & \Co{z}{6}{G}{o}{7/2} &  8  &  3.25 &   18.3  &   6.~~~& 95 &
\Co{f}{4}{F}{ }{5/2} &  6  &  6.04 &   19.5  &  3.~~ \\ 
 46 & \Co{z}{6}{G}{o}{5/2} &  6  &  3.28 &   14.4  &   3.~~~& 96 &
\Co{f}{4}{F}{ }{3/2} &  4  &  6.09 &  --1.5  &  0.~~ \\ 
 47 & \Co{z}{6}{G}{o}{3/2} &  4  &  3.30 &    4.9  &   4.~~~& 97 &
\Co{g}{4}{F}{ }{9/2} & 10  &  6.34 &    9.2  &  0.~~ \\ 
 48 & \Co{z}{4}{F}{o}{9/2} &  10  &  3.51 &   27.~~~& --2.~~~& 98 &
\Co{g}{4}{F}{ }{7/2} &  8  &  6.35 &   11.2  &  0.~~ \\
 49 & \Co{z}{4}{F}{o}{7/2} &  8  &  3.57 &   21.9  &   1.~~~& 99 &
\Co{g}{4}{F}{ }{5/2} &  6  &  6.46 &   13.9  &  0.~~ \\
 50 & \Co{z}{4}{F}{o}{5/2} &  6  &  3.62 &   18.7  &   1.~~~&100 &
\Co{g}{4}{F}{ }{3/2} &  4  &  6.53 &   14.~~~&  0.~~ \\
\ion{Co}{ii} & & & & & & & & & & & \\
 1   &   \Co{a}{5}{P}{}{3} &  7  & 2.20 &  40.0  & --20.0  & 4 &
\Co{z}{5}{D}{o}{4} &  9  & 5.74 &   8.0  & 10.0 \\
 2   &   \Co{a}{5}{P}{}{2} &  5  & 2.23 &  49.9  &   20.0  & 5 &
\Co{z}{5}{D}{o}{3} &  7  & 5.83 &   8.0 & 0.~~ \\
 3   &   \Co{a}{5}{P}{}{1} &  3  & 2.27 &  60.0  &    2.0  & 6 &
\Co{z}{5}{D}{o}{2} &  5  & 5.89 & --30.0 & 12.0 \\
\noalign{\smallskip}\hline\noalign{\smallskip}
\end{tabular}
\end{center}
\end{minipage}
\end{table*}

\bsp

\label{lastpage}
\end{document}